\documentclass[sigconf, 10pt, screen]{acmart}

\usepackage[ruled,linesnumbered]{algorithm2e}
\usepackage{makecell}
\usepackage{titlesec}
\usepackage[subtle]{savetrees}
\usepackage{enumitem}
\usepackage{bm}
\usepackage{breqn}
\usepackage{hyperref}

\graphicspath{{figs/}}
\def\BibTeX{{\rm B\kern-.05em{\sc i\kern-.025em b}\kern-.08em
    T\kern-.1667em\lower.7ex\hbox{E}\kern-.125emX}}

\newcommand{\figwidth}{.7\linewidth}

\titlespacing*{\section}{0pt}{2pt}{2pt}
\titlespacing*{\subsection}{0pt}{2pt}{2pt}
\titlespacing*{\subsubsection}{0pt}{4pt}{4pt}
\setlist{nosep}
\copyrightyear{2025} 
\acmYear{2025} 
\setcopyright{rightsretained}
\acmConference[ICS '25]{2025 International Conference on Supercomputing}{June 8--11, 2025}{Salt Lake City, UT, USA}
\acmBooktitle{2025 International Conference on Supercomputing (ICS '25), June 8--11, 2025, Salt Lake City, UT, USA}
\acmDOI{10.1145/3721145.3725773}
\acmISBN{979-8-4007-1537-2/2025/06}

\begin{document}

\title{MAGNUS: Generating Data Locality to Accelerate Sparse Matrix-Matrix Multiplication on CPUs}

\author{Jordi Wolfson-Pou}
\affiliation{
  \institution{Intel Labs}
  \city{Santa Clara}
  \state{California}
  \country{USA}
}
\email{jordi.wolfson-pou@intel.com}

\author{Jan Laukemann}
\affiliation{
  \institution{Friedrich-Alexander-Universit{\"a}t Erlangen-N{\"u}rnberg, Erlangen National High Performance Computing Center}
  \city{Erlangen}
  \country{Germany}
}
\email{jan.laukemann@fau.de}

\author{Fabrizio Petrini}
\affiliation{
  \institution{Intel Labs}
  \city{Santa Clara}
  \state{California}
  \country{USA}
}
\email{fabrizio.petrini@intel.com}

\begin{abstract}
Sparse general matrix-matrix multiplication (SpGEMM) is a critical operation in many applications.  Current multithreaded implementations are based on Gustavson’s algorithm and often perform poorly on large matrices due to limited cache reuse by the accumulators.  We present MAGNUS (Matrix Algebra for Gigantic NUmerical Systems), a novel algorithm to maximize data locality in SpGEMM.  To generate locality, MAGNUS reorders the intermediate product into discrete cache-friendly chunks using a two-level hierarchical approach.  The accumulator is applied to each chunk, where the chunk size is chosen such that the accumulator is cache-efficient.  MAGNUS is input- and system-aware: based on the matrix characteristics and target system specifications, the optimal number of chunks is computed by minimizing the storage cost of the necessary data structures.  MAGNUS allows for a hybrid accumulation strategy in which each chunk uses a different accumulator based on an input threshold.  We consider two accumulators: an AVX-512 vectorized bitonic sorting algorithm and classical dense accumulation.  An OpenMP implementation of MAGNUS is compared with several baselines, including Intel MKL, for a variety of different matrices on three Intel architectures.  For matrices from the SuiteSparse collection, MAGNUS is faster than all the baselines in most cases and is often an order of magnitude faster than at least one baseline.  For massive random matrices, MAGNUS scales to the largest matrix sizes, while the baselines do not.  Furthermore, MAGNUS is close to the optimal bound for these matrices, regardless of the matrix size, structure, and density.
\end{abstract}



\maketitle

\section{Introduction}
\sloppy The sparse general matrix-matrix multiplication operation (SpGEMM) $C = A B$ is critical to the performance of many applications, including genome assembly~\cite{genome1,genome2,bella}, machine learning~\cite{pca,mcl,dnn,HipMCL,hoefler}, algebraic multigrid~\cite{amg1,amg2}, and graph analytics~\cite{bfs,subgraph,tricount1,tricount2,color,tricount3}.
The main challenge of SpGEMM comes from the sparse structures of $A$ and $B$, leading to unpredictable memory access patterns. Such irregularities pose significant difficulties for modern multicore architectures optimized for regular access patterns and high data reuse. Consequently, many state-of-the-art SpGEMM algorithms struggle to scale effectively for massive, irregular matrices, mainly due to inefficient utilization of the cache hierarchy.

Multithreaded SpGEMM algorithms are typically based on Gustavson's method, where, for each row of $A$, rows of $B$ are loaded, scaled, \emph{accumulated}, and written to $C$. The performance of the accumulation step is critical to the overall performance of SpGEMM. Conventional accumulators perform well for certain sparsity patterns, such as banded matrices, where the entire accumulator data structure is accessed infrequently. However, for highly irregular matrices, such as random power-law matrices, frequent accesses to the entire accumulator lead to suboptimal data reuse. As a result, scaling to massive matrices can become prohibitive, especially when the size of the accumulator exceeds the highest level of private cache.
Several algorithms have been proposed to address caching issues in accumulators, the most recent being the CSeg method~\cite{partway,cseg}, where $B$ is partitioned into segments so that only a smaller range of the accumulator is accessed. However, CSeg scales poorly for some datasets; when there are many partitions, the cost of constructing and accessing the partitioning information becomes significant, especially as the number of partitions increases with the matrix dimensions.

We present MAGNUS (\textbf{M}atrix \textbf{A}lgebra for \textbf{G}igantic \textbf{NU}merical \textbf{S}ystems), a novel algorithm that uses a hierarchical approach to generate the locality needed by the accumulators.
The central idea of MAGNUS is to reorder the \textit{intermediate product} of $C$ (arrays of column indices and values generated before accumulation) into cache-friendly chunks that are processed independently by the accumulator.
The MAGNUS workflow consists of two main algorithms: the fine- and coarse-level algorithms, the naming of which comes from two-level multigrid methods~\cite{amg1}.
The coarse-level algorithm is based on the outer product formulation of SpGEMM and generates the first level of locality.
The fine-level algorithm is based on Gustavson's formulation and further reorders the coarse-level chunks. The accumulator is then applied to each fine-level chunk.
MAGNUS is input- and system-aware: the number of fine- and coarse-level chunks are chosen based on the matrix parameters and system specifications,
where the optimal number of chunks is selected by minimizing the storage requirement of all frequently accessed data structures.
Additionally, MAGNUS is accumulator agnostic, where conventional accumulators can be applied to the fine-level chunks.
This paper considers two accumulators: AVX-512 vectorized sorting, which is used on chunks with a small number of elements, and dense accumulation, which is used otherwise.

Our experimental results provide two significant contributions: a set of microbenchmarks to motivate the need for MAGNUS, and the comparison of an OpenMP implementation of MAGNUS with six state-of-the-art SpGEMM baselines.
For the microbenchmarks, the key building blocks of MAGNUS are tested in isolation and analyzed using Likwid~\cite{likwid}.
First, we show that the performance of the accumulators drops significantly if the accumulation data structures do not fit into the L2 cache. Second, we show that with the optimal MAGNUS parameters, the execution time for the building blocks is minimized and performs at near-streaming speed.

For the SpGEMM results, MAGNUS is compared to six state-of-the-art baselines, including
CSeg~\cite{cseg} and Intel Math Kernel Library (MKL)~\cite{mkl}.
Three matrix test sets are evaluated on three Intel architectures.
For the SuiteSparse matrix collection~\cite{suitesparse}, MAGNUS is the fastest method in most cases and is often an order of magnitude faster than at least one baseline.
For our second matrix set, which comes from a recursive model to generate power law graphs~\cite{rmat}, MAGNUS is always the fastest.
With the exception of CSeg, the speedup of MAGNUS over the baselines increases as the matrix size increases.
Lastly, we consider massive uniform random matrices~\cite{erdosrenyi}, which is the most challenging case for our baselines since the uniformity results in frequent accesses to the entire accumulation data structures.
This test set demonstrates the need for the two-level approach in MAGNUS, where using the fine-level algorithm alone results in divergence from an ideal performance bound (CSeg also exhibits this poor scaling).
However, using the complete MAGNUS algorithm allows scaling to the largest case, where the performance of MAGNUS is close to an ideal bound independent of the matrix size.
\section{Background}\label{sec:background}
\subsection{Gustavson's Method}\label{sec:gustav}
For a sparse matrix $X$, we define $n_X$, $m_X$, and $nnz_X$ as the number of rows, columns, and nonzero entries, respectively. The set $\mathcal{S}(X)$ denotes the column indices of all nonzero entries in $X$.
Our notation also extends to individual rows.
For example, for row $i$ of $X$, $nnz_{X_i}$ denotes the number of nonzero entries, and $\mathcal{S}(X_i)$ denotes the set of column indices corresponding to the nonzero entries.

The general sparse matrix-matrix multiplication operation (SpGEMM) is defined as $C=A B$, where $A$, $B$, and $C$ are sparse matrices.
Multithreaded implementations of SpGEMM are typically based on Gustavson's row-by-row algorithm~\cite{gustavson}:
\begin{equation}
    C_i = \sum_{j\in \mathcal{S}(A_i)} A_{ij}B_j,
    \label{equ:gustav}
\end{equation}
i.e., for some row $i$ of $C$, the rows of $B$ are scaled by the nonzero values in row $i$ of $A$.
These scaled rows are then summed together to give the final row of $C$.
Since each row of $C$ is computed independently, multithreaded implementations typically partition rows among threads,
which is the approach we take for MAGNUS.

There are two main ingredients for implementing Gustavson's method: the matrix storage scheme and the algorithm that \emph{accumulates} the scaled rows of $B$.
Compressed sparse row (CSR) format is one of the most popular storage schemes and is especially useful for algorithms such as Gustavson's method that traverse matrices row-wise. 
The CSR format requires three arrays: $C.col$, $C.val$, and $C.rowPtr$.
Arrays $C.col$ and $C.val$ of size $nnz_C$ store the column indices and values, respectively, of the nonzero entries of $C$, and $C.rowPtr$, of size $n_C+1$, stores the starting positions of each row in $C.col$ and $C.val$.

\autoref{alg:gustav_dense} shows the pseudocode for the \emph{numeric phase} of Gustavson's method. Variables in \textbf{bold} are global, meaning they are shared and visible across all threads.  
The scaled rows of $B$ are summed using a \emph{dense accumulator}, defined as the combination of \texttt{denseAccumBuff} and \texttt{bitMap}.
The column indices in row $i$ of $A$ are loaded as $idx = A.col[A.rowPtr[i]]$, and the rows of $B$ are loaded by reading $B.col$ from $B.rowPtr[idx]$ to $B.rowPtr[idx+1]$.
Array $denseAccumBuff$ of size $m_C$ is updated
for each column index of $B$, a companion bitmap stores the nonzero positions in $denseAccumBuff$, and $colBuff$ stores the running list of column indices in $C$.
Besides $A$, $B$, and $C$, all variables are thread-local.

\autoref{alg:gustav_dense} can be extended to other types of accumulators, e.g., hash map-based accumulators, where the dense accumulation array and bitmap are replaced by a hash map.
In other Gustavson-based algorithms, such as expand-sort-compress (ESC)~\cite{ESC,pbSpGEMM}, the \emph{intermediate product} of $C$ is written to an array instead of directly updating the accumulator.
This is shown in \autoref{alg:gustav_sort},
where the intermediate product is generated in the first loop by storing $B.col[k]$ and $A.val[j] \times  B.val[k]$ in 
$colBuff$ and $valBuff$, respectively.
The intermediate product is then sorted, duplicates are merged, and the result is written to $C$ (these steps take place in \texttt{sortMergeWrite(}$colBuff$,$valBuff$\texttt{)}).

\begin{algorithm}[htbp]
    \small
    \caption{Gustavson SpGEMM: Dense Accumulation}\label{alg:gustav_dense}
    \KwIn{$\bm{A}$, $\bm{B}$, $\bm{C.rowPtr}$}
    \KwOut{$\bm{C.col}$, $\bm{C.val}$}
    \SetCommentSty{emph}
    \DontPrintSemicolon
    \SetKwFor{ForPar}{for}{do \emph{in parallel}}{end forpar}
    \ForPar{$i \gets 0$\textbf{ to }$n-1$}{
        $count\gets 0$, $denseAccumBuff\gets 0$\;
        \text{\textbf{\color{blue}/* Read row $i$ of $A$ */}}\;
        \For{$j \gets \boldsymbol{A.rowPtr}[i]$\textbf{ to }$\boldsymbol{A.rowPtr}[i+1]-1$}{
            \text{\textbf{\color{blue}/* Read row $j$ of $B$ */}}\;
            \For{$k \gets \boldsymbol{B.rowPtr}[\boldsymbol{A.col}[j]]$\textbf{ to }$\boldsymbol{B.rowPtr}[\boldsymbol{A.col}[j]+1]-1$}{
                \text{\textbf{\color{blue}/* Multiply and update accumulator */}}\;
                $denseAccumBuff[\boldsymbol{B.col}[k]]$ \texttt{+=} $\boldsymbol{A.val}[j] \times  \boldsymbol{B.val}[k]$\;
                \If{$bitMap[\boldsymbol{B.col}[k]]$ \texttt{==} $0$}{
                    $colBuff[count\text{\texttt{++}}] \gets \boldsymbol{B.col}[k]$\;
                    $bitMap[\boldsymbol{B.col}[k]] \gets 1$\;
                }
            }
        }
        \text{\textbf{\color{blue}/* Write to $C$ */}}\;
        $k \gets \boldsymbol{C.rowPtr}[i]$\;
        \For{$j \in colBuff$}{
            $\boldsymbol{C.col}[k] \gets j$\;
            $\boldsymbol{C.val}[k\texttt{++}] \gets denseAccumBuff[j]$\;
            $bitMap[j] \gets 0$\;
        }
    }
\end{algorithm}

\begin{algorithm}[htbp]
    \small
    \caption{Gustavson SpGEMM: Expand-Sort-Compress (ESC)}\label{alg:gustav_sort}
    \KwIn{$\bm{A}$, $\bm{B}$, $\bm{C.rowPtr}$}
    \KwOut{$\bm{C.col}$, $\bm{C.val}$}
    \SetCommentSty{emph}
    \DontPrintSemicolon
    \SetKwFor{ForPar}{for}{do \emph{in parallel}}{end forpar}
    \ForPar{$i \gets 0$\textbf{ to }$n-1$}{
        $count\gets 0$\;
        \text{\textbf{\color{blue}/* Read row $i$ of $A$ */}}\;
        \For{$j \gets \boldsymbol{A.rowPtr}[i]$\textbf{ to }$\boldsymbol{A.rowPtr}[i+1]-1$}{
            \text{\textbf{\color{blue}/* Read row $j$ of $B$ */}}\;
            \For{$k \gets \boldsymbol{B.rowPtr}[\boldsymbol{A.col}[j]]$\textbf{ to }$\boldsymbol{B.rowPtr}[\boldsymbol{A.col}[j]+1]-1$}{
                \text{\textbf{\color{blue}/* Multiply and update accumulator */}}\;
               $colBuff[count] \gets \boldsymbol{B.col}[k]$\;
               $valBuff[count\text{\texttt{++}}] \gets \boldsymbol{A.val}[j] \times  \boldsymbol{B.val}[k]$\;
            }
        }
        \text{\textbf{\color{blue}/* Sort, merge, and write to $C$ */}}\;
        \texttt{sortMergeWrite(}$colBuff$,$valBuff$\texttt{)}\;
    }
\end{algorithm}

To compute $C.rowPtr$, which is an input to the numeric phase, an initial \emph{symbolic phase} is required.
The symbolic phase typically has the same high-level algorithm as the numeric phase but without performing the multiplication $A.val[j] \times  B.val[k]$ and writing to $C$.
For example, in \autoref{alg:gustav_dense}, the symbolic phase does not include the modifications to $denseAccumBuff$, $C.col$, and $C.val$.
Instead, only the bitmap is updated along with a counter that outputs the exact number of nonzero entries for each row of $C$.
Finally, $C.rowPtr$ is computed using a prefix sum on the counters.
This type of symbolic phase is known as \textit{precise prediction}, where the number of nonzero entries in $C$ is calculated exactly before the numeric phase.

\sloppy On modern CPUs, maximizing cache reuse is crucial to the performance of any application.
In SpGEMM, the accumulator is the most frequently accessed, where the amount of reuse is determined by the sparsity pattern of $A$ and $B$.
For optimal performance, the dense accumulator should be confined to the L2 cache, which is the highest level of private cache.
This efficient cache utilization occurs naturally in specific matrix structures, such as banded matrices or matrices that yield a highly sparse $C$. However, for matrices that produce ``large'' intermediate products (where ``large'' refers to both the number of nonzero elements and a wide distribution of column index values), SpGEMM faces significant challenges. A prominent example is random power-law matrices that model social networks~\cite{rmat}. For such matrices, the size of $denseAccumBuff$ often exceeds the capacity of the L2 cache and the large intermediate product results in frequent accesses to the entire $denseAccumBuff$ array. Consequently, $denseAccumBuff$ must frequently be evicted from and reloaded to the L2 cache, resulting in suboptimal performance. This breakdown in locality presents a substantial obstacle for current SpGEMM algorithms, as we will demonstrate through microbenchmarks and SpGEMM results.

\begin{figure*}[th]
\centering
\begin{tabular}{c}
\includegraphics[width=\linewidth]{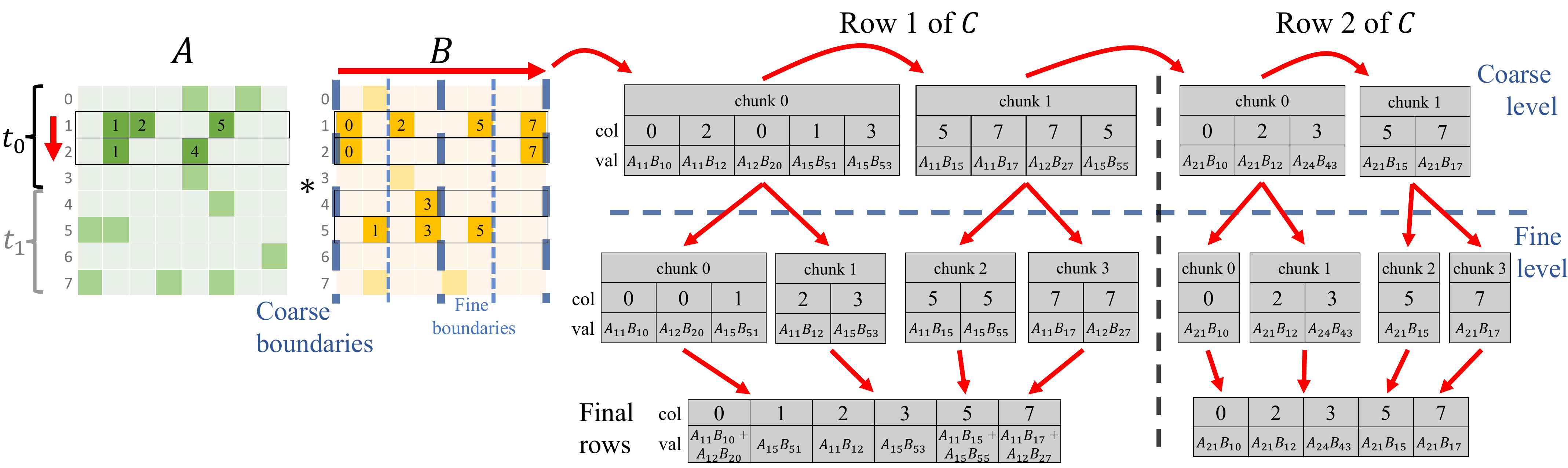} 
\end{tabular}
\caption{Example workflow of MAGNUS, where two threads multiply two $8\times8$ matrices.
The computation of rows 1 and 2 assigned to thread $t_0$ is shown.
Two coarse- and fine-level chunks are used for each row.
}
\label{fig:MAGNUS_example}
\end{figure*}

\subsection{Related Work}
Optimizations related to the accumulation step have been the primary focus of research on SpGEMM.
For load balancing, a common approach is based on the observation that rows with different intermediate sizes require different accumulation strategies~\cite{liu1,liu2,ESC,pbSpGEMM,liu3,OpSparse,spECK,nagasaka3}.
In~\cite{ESC}, $A$ is reordered by increasing intermediate size.
Other approaches group rows based on the size of the intermediate product, where a different accumulator is used for different groups~\cite{liu1,liu2,liu3,OpSparse,spECK,nagasaka3,adaptLoadBalanceGPU}.
In~\cite{liu2}, five different accumulation strategies are used, including priority queues and sorting algorithms.
Developing new accumulators is another topic that has been widely studied, especially for modern multicore machines with vector processors~\cite{fevre,hierRowMerging,ASA,AVX512sort,nagasaka1,nagasaka2,gamma}.
A common approach is to optimize sorting algorithms~\cite{ESC,tricount3,liu1,AVX512sort,pbSpGEMM,registerAware}
or data structures such as heaps~\cite{azad,nagasaka1,nagasaka2} and hash maps~\cite{nagasaka1,nagasaka2,cuSPARSE,registerAware,mcl,balancedHashing,kokkos}.
In \autoref{sec:results_spgemm}, MAGNUS is compared with hash map and heap-based approaches, which are considered state-of-the-art~\cite{cseg,nagasaka1,nagasaka2,mkl}.

Perhaps most relevant to MAGNUS are recent works on improving the cache behavior of accumulators,
proposed in~\cite{partway} and improved in the CSeg method~\cite{cseg}.
The core concept is to partition the columns of $B$ into segments, where the number of segments is chosen so that the dense accumulator fits in the L2 cache.
An additional \emph{high-level summary matrix} is used to store the segmentation information. 
CSeg was shown to be overall faster than many state-of-the-art libraries mentioned above.
For a more extensive overview of SpGEMM research, including distributed memory algorithms, see~\cite{spgemmSurvey}.
\section{MAGNUS}\label{sec:magnus}
\subsection{Overview}
MAGNUS uses a novel hierarchical algorithm to generate two levels of locality.
The \emph{coarse-level algorithm} reorders the intermediate product into discrete chunks
and the \emph{fine-level algorithm} further subdivides and accumulates the coarse-level chunks.
The number of chunks used in both algorithms is based on optimal parameters that are computed using the
input matrix properties and the target system specifications.
These parameters are optimal in the sense that they minimize the storage cost of all frequently accessed arrays.
The levels are generated using a set of basic operations, including histogramming and prefix summing.
Combining the building blocks and the optimal parameters creates the locality required by the accumulators.

For sufficiently ``small'' matrices (as discussed later in the derivation of the MAGNUS parameters), the fine-level algorithm alone provides an adequate level of data locality. This algorithm is based on Gustavson's method, similar to most SpGEMM algorithms. However, achieving scalability for ``massive'' matrices requires both fine- and coarse-level locality. Here, massive matrices are those in which the data structures required by the fine-level algorithm, including the accumulator, exceed the capacity of the L2 cache. The coarse-level algorithm employs an outer product-based approach, necessitating an additional pass over the intermediate product, which increases the total data volume.
This additional cost means that using the standalone fine-level algorithm wherever possible is advantageous, which is why we reserve the coarse-level algorithm only for massive matrices.
Since the intermediate product is generated in both approaches, MAGNUS can be classified as an ESC-type algorithm~\cite{ESC}.

\autoref{fig:MAGNUS_example} shows a simple workflow for MAGNUS, where two threads are used to multiply two $8\times 8$ matrices.
The example shows how rows 1 and 2 assigned to thread $t_0$ are computed, where two chunks are used for both the coarse and fine levels.
The outer product-based approach traverses the submatrix of $A$ (corresponding to rows 1 and 2) column by column, and the highlighted rows of $B$ are traversed row by row.
This traversal generates the four coarse-level chunks.
Each chunk is reordered again to get the eight fine-level chunks, where the accumulator is applied to get the final result.
It is important to note that all coarse-level chunks are generated before executing the fine-level algorithm.
However, the fine level is processed depth-first: for each coarse-level chunk, the fine level is generated and then immediately accumulated (similar to a Gustavson-based method) before proceeding to the next coarse-level chunk.

The key property of MAGNUS is that the range of column indices in each fine-level chunk is significantly smaller than $m_C$ (the number of columns of $C$).
This allows the dense accumulator to fit in the L2 cache when $m_C$ exceeds the L2 cache capacity. In \autoref{fig:MAGNUS_example}, the per-chunk range of column indices is two, compared to $m_C=8$. For example, the column indices in the fourth fine-level chunk of either row fall within the range $[6,7]$.
Although not shown in the figure, the column indices in each chunk are shifted into the chunk-local range $[0,1]$, which means that we only need a dense accumulator of length two.
If we consider a theoretical system with an L2 cache capable of storing a dense accumulator with a maximum of two elements, each fine-level chunk can be accumulated with minimal L2 cache misses.
In contrast, if we used Gustavson's method with a conventional dense accumulator, L2 cache misses would occur frequently after loading the first two elements in the second row of $B$. 

The high-level steps of MAGNUS are: (1) the setup phase, which includes calculating the optimal number of chunks (this will be discussed later in \autoref{sec:magnus_opt_params}), computing the intermediate product size for each row, and categorizing each row; (2) the symbolic phase; and (3) the numeric phase.
The setup phase is inexpensive compared to the symbolic and numeric phases: calculating the optimal number of chunks has constant time complexity, 
while the remaining setup steps involve a highly parallel single pass over the rows of $C$, with time complexity $O(n_C/t)$, where $t$ is the number of threads.
Row categorization is necessary because not all rows require locality generation. 

MAGNUS categorizes each row based on its structure and the system's specifications:
\begin{enumerate}
    \item \textbf{Sort}: If the number of intermediate elements is less than the dense accumulation threshold, we can directly apply a sort-based accumulator, as in \autoref{alg:gustav_sort}.  The dense accumulation threshold will be described later in this section.
    \item \textbf{Dense accumulation}: If the \emph{intermediate row length} fits into the L2 cache, we can directly apply dense accumulation to the row, as in \autoref{alg:gustav_sort}.  This is because the range of column indices does not exceed the size of the L2 cache.   The intermediate row length is the difference between the minimum and maximum column index of the intermediate product.
    \item \textbf{Fine level:} If $s_{finelevel} < s_{L2}$, the fine-level algorithm can be applied, where $s_{finelevel}$ is the number of bytes required to store all necessary fine-level data structures (this is discussed later in this section).
    \item \textbf{Coarse level:} The coarse-level algorithm is applied to all remaining rows, where the fine-level algorithm is applied to each coarse-level chunk.
\end{enumerate}
The accumulation parameters mentioned above will be discussed in \autoref{sec:magnus_accum}, including a description of the sorting algorithm.
The first two categories imply that some rows possess intrinsic locality and do not benefit from the locality-generation algorithms in MAGNUS.
After the rows are categorized, the rows of $C$ are computed category-first to ensure that data specific to a particular category are cached for as long as possible.
In our OpenMP implementation of MAGNUS, we use a parallel for loop with dynamic scheduling to traverse the rows in each category with a \texttt{no wait} clause.
The \texttt{no wait} clause ensures that threads proceed to the next category without unnecessary synchronization.

For simplicity, we assume $m_C$ is a power of two in our descriptions of the algorithms in MAGNUS.
We use precise prediction for the symbolic phase, but for brevity, we will only describe the numeric phase.
See \autoref{sec:background} for clarity on the differences between the symbolic and numeric phases.

\subsection{The Fine-level Algorithm}\label{sec:magnus_fineLevel}
The fine-level algorithm has the following steps for each row: histogram, prefix sum, reorder, and accumulate, where $n_{chunksFine}$ is the number of fine-level chunks.
As in Gustavson's method, each row (or coarse-level chunk in cases where the coarse-level algorithm is applied) is computed before moving on to the next row.
This means that the intermediate product is generated only for a single row (or coarse-level chunk) at any given time, unlike
in outer-product-based approaches.

Pseudocode for the fine-level algorithm is shown in \autoref{alg:magnus_fine}, where the input is the column indices and values of a single coarse-level chunk for row $i$ of $C$.
For rows that only require fine-level locality, $A$ and $B$ are read directly as in \autoref{alg:gustav_dense}, i.e.,
the loop headers on lines 3 and 12 of \autoref{alg:magnus_fine} are replaced with the nested loop headers on lines 4 and 6 of \autoref{alg:gustav_dense}, respectively.
The notation $arr \gets \alpha$ means that the entire array $arr$ is initialized to the value $\alpha$.
The C-style notation $arr[i]$ means that element $i$ of $arr$ is accessed, and $\&arr[i]$ means that the array is accessed starting at element $i$.

\begin{algorithm}[htbp]
    \small
    \SetCommentSty{emph}
    \DontPrintSemicolon
    \caption{MAGNUS fine-level algorithm applied to a single coarse-level chunk}\label{alg:magnus_fine}
    \KwIn{$colCoarse$, $valCoarse$}
    \KwOut{$\boldsymbol{C}_{i,rangeCoarse}$}
    
    \SetKwFor{ForPar}{for}{do \emph{in parallel}}{end forpar}
    \text{\textbf{\color{blue}/* Histogram */}}\;
    $countsFine\gets 0$\;
    \For{$col \in colCoarse$}{
        $chunk\gets col$ \texttt{>>} $shiftFine$\;
        $countsFine[chunk]$\texttt{++}\;
    }
    \text{\textbf{\color{blue}/* Prefix sum */}}\;
    $offsetsFine[0] \gets 0$\;
    $\&offsetsFine[1] \gets$ \texttt{inclusiveScan(}$countsFine$\texttt{)}\;
    \text{\textbf{\color{blue}/* Reorder */}}\;
    $countsFine\gets 0$\;
    \For{$\{col, val\} \in \{colCoarse, valCoarse\}$}{
        $chunk\gets col$ \texttt{>>} $shiftFine$\;
        $\ell \gets offsetsFine[chunk]+countsFine[chunk]$\texttt{++}\;
        $colFine[\ell]\gets col - chunk\times chunkLenFine$\;
        $valFine[\ell]\gets val$\;
    }
    \text{\textbf{\color{blue}/* Accumulation */}}\;
    \For{$j \gets 0$\textbf{ to }$n_{chunksFine}-1$}{
        $k \gets offsetsFine[j]$\;
        $\boldsymbol{C}_{i,rangeFine_j} \gets$ \texttt{accum(}$\&colFine[k],\&valFine[k]$\texttt{)}\;
    }
\end{algorithm}

\begin{figure}[htbp]
\centering
\begin{tabular}{c}
\includegraphics[width=0.61\linewidth]{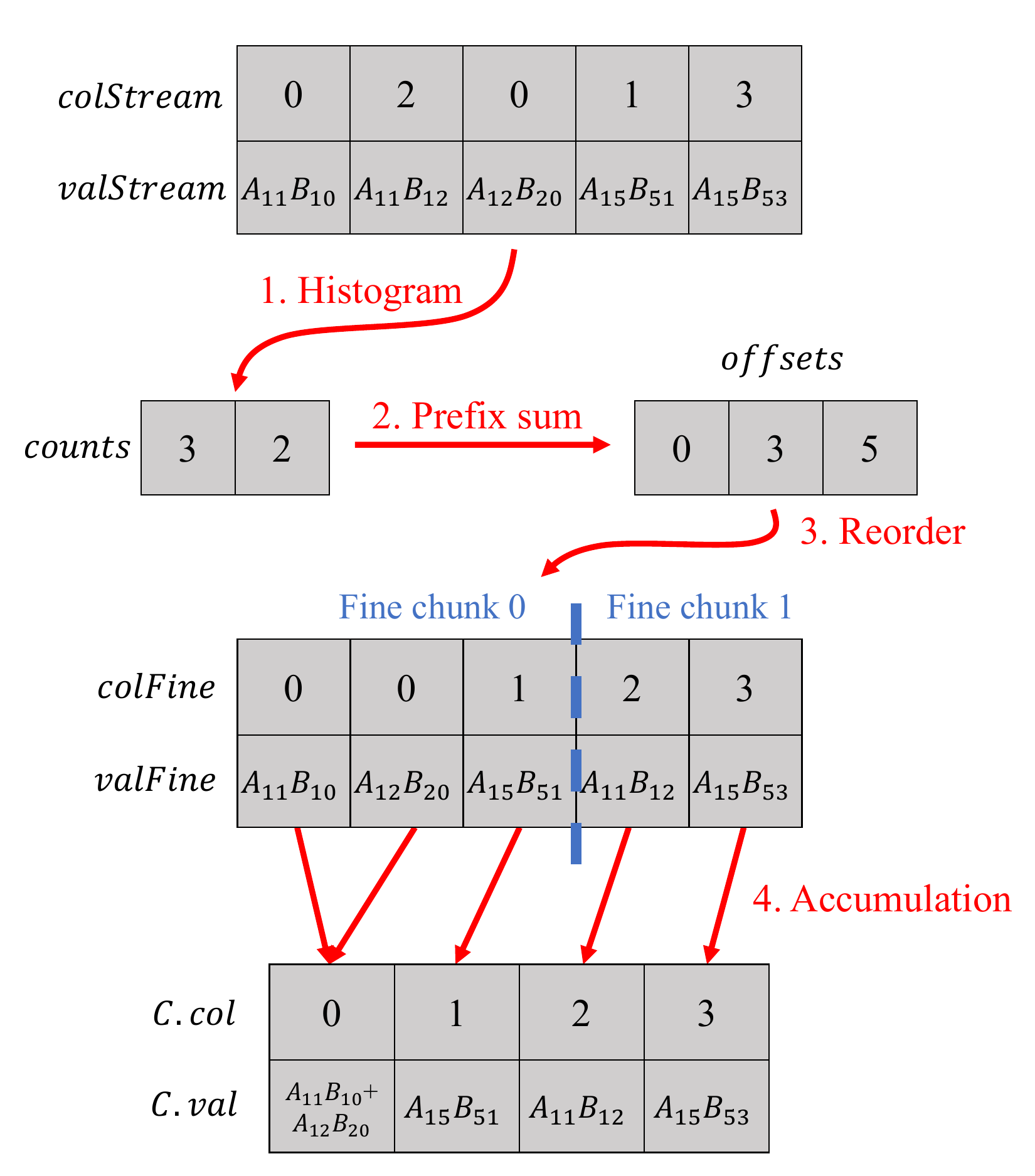} 
\end{tabular}
\caption{Data structure-view of applying the fine-level algorithm to the first chunk from \autoref{fig:MAGNUS_example}.}
\label{fig:fineLevel_example}
\end{figure}

The first step towards reordering the intermediate product is to compute $offsetsFine$ (the chunk offsets) using a histogram and a prefix sum operation.  
The array $offsetsFine$ stores the start and end locations of each fine-level chunk in $colFine$ and $valFine$.
The buffers $colFine$ and $valFine$ are typical of any ESC-type algorithm where the intermediate product must be explicitly stored.  In the case of MAGNUS, they store the reordered intermediate product.
In the histogram step, the column indices are mapped to chunks as $col / chunkLenFine$, where $chunkLenFine = m_{C_{maxL2}} / n_{chunksFine}$ is the \textit{chunk length} of the fine-level chunks.
The chunk length is the local range of the column indices within a chunk, where column indices are shifted into the range $[0,chunkLenFine)$ as shown on line 15.
The value of $m_{C_{maxL2}}$ is equal to $m_C$ if the fine-level algorithm is used alone, or equal to the coarse chunk length if the coarse-level algorithm is used ($m_{C_{maxL2}}$ is discussed in more detail in \autoref{sec:magnus_opt_params}).
To optimize the mapping, the division operation is replaced with a bitwise operation by restricting $n_{chunksFine}$ to a power of two, ensuring that $m_C / n_{chunksFine}$ is also a power of two.
Therefore, the mapping becomes $col \texttt{>>} shiftFine$, where $shiftFine = \log_2(chunkLenFine)$ and $\texttt{>>}$ is the bitwise right-shift operator.

After the histogram is computed, the chunk offsets are computed using a prefix sum (inclusive scan) of the histogram.
To reorder the input (shown in the second loop), the histogramming phase is repeated, where the histogram is used to track the current
number of elements in each chunk.
Elements are reordered by writing the input coarse-level chunk at the position $offsetsFine[chunk]+countsFine[chunk]$ in $colFine$ and $valFine$.
As mentioned previously, the column indices are shifted into the local range of each chunk as $col - chunk\times chunkLenFine$ to allow for cache-efficient accumulation.

Finally, a call to \texttt{accum()} for each chunk invokes either sort-based accumulation or dense accumulation, where the size of $denseAccumBuff$ (see \autoref{alg:gustav_dense}) is now reduced from $m_{C_{maxL2}}$ to $chunkLenFine$.
After the accumulation step, the column indices are shifted back into the correct range before writing to $C$.
The variable $rangeFine_j$ denotes the range of column indices of the fine-level chunk $j$ (i.e., $[j\times chunkLenFine,(j+1)\times chunkLenFine)$), and $rangeCoarse$
is the range of column indices of the input coarse-level chunk.
\autoref{fig:fineLevel_example} shows the workflow of the fine-level algorithm in terms of the data structures from \autoref{alg:magnus_fine}, where the input is the first chunk from the example in \autoref{fig:MAGNUS_example}.

\sloppy The fine-level algorithm requires two additional arrays: $countsFine$ and $offsetsFine$, both of size $n_{chunksFine}$. Alongside $denseAccumBuff$ and $bitMap$, the goal is to keep $countsFine$, $offsetsFine$, and the active cache lines of $colFine$ and $valFine$ in the L2 cache.
The active cache lines must be considered since we are writing to $colFine$ and $valFine$ at noncontiguous positions.
In our implementations, we use non-temporal streaming stores when writing to $colFine$ and $valFine$, which avoids polluting the L2 cache.
Non-temporal stores are intrinsic functions used on Intel processors (e.g., \texttt{_mm512\_stream\_si512()}) that write to memory without evicting cached data, allowing us to retain the accumulator and fine-level data structures in the L2 cache while streaming the intermediate product.
\autoref{sec:magnus_opt_params} shows how we choose the optimal number of chunks that minimizes the total storage cost of these cached arrays.

\subsection{The Coarse-level Algorithm}\label{sec:magnus_coarse}
As the columns of $C$ increase, the storage of the fine-level data structures eventually exceeds the size of the L2 cache.
An initial coarse level must be generated for such matrices, providing the first level of locality.
To generate the coarse level, we use a modified outer product-based algorithm, where the intermediate product is generated and reordered for all rows that require coarse-level locality before any accumulation occurs.
The reordered intermediate product is organized into discrete coarse-level chunks that can be handled independently by the fine-level algorithm.

The outer product-based approach is used because the coarse-level algorithm computes only the intermediate product without performing any accumulation.
Therefore, maximizing the reuse of the input matrices rather than the accumulator is beneficial, which is a well-known property of outer product-based SpGEMM algorithms~\cite{pbSpGEMM,gamma}.
The combination of our reordering algorithm with the outer product formulation makes the coarse-level algorithm similar to propagation blocking-based algorithms~\cite{pbSpGEMM}.

Conventional outer product algorithms typically generate the intermediate product by multiplying the columns of $A$, stored in CSC format, with the rows of $B$, stored in CSR format.
The coarse-level algorithm in MAGNUS follows the same scheme, but on the subset of the rows categorized as coarse-level rows.
Therefore, a CSC version of the submatrix $\hat{A}$ of $A$ is constructed, where $\hat{A}$ only includes these coarse-level rows.
Each thread performs the following steps on its list of coarse-level rows, which are stored in the array $coarseRowsC$:
\begin{enumerate}
    \item For all $i\in coarseRowsC$, generate $coarseRowsB$, i.e., the unique set of rows in $B$ required to perform the outer product.  This is done by iterating through the nonzero entries in $\hat{A}$ and setting a bitmap, where $coarseRowsB$ is the list of set bits.
    \item Construct the thread-local CSC submatrix $\hat{A}^{CSC}$ using the well-known approach for the conversion of CSR to CSC: a histogramming stage computes the number of nonzero entries per row, a prefix sum computes $\hat{A}^{CSC}.colPtr$, and $\hat{A}^{CSC}.colPtr$ is then used to construct the arrays $\hat{A}^{CSC}.row$ and $\hat{A}^{CSC}.val$.
    This step is inexpensive since the time complexity of outer product-based SpGEMM algorithms is dominated by the generation of the intermediate product in the following step. Specifically, the time complexity of constructing $\hat{A}$ is $O(nnz_{\hat{A}})$, while generating the intermediate product requires $O(\sum_{(i,j)\in \mathcal{S}(\hat{A})}nnz_{B_j})$ time, which is best-case $O(nnz_{\hat{A}})$ (i.e., each row in $B$ contains only one nonzero value) and worst-case $O(nnz_{\hat{A}} m_B)$ (i.e., each row in $B$ is dense).
    \item Perform a histogram step by reading $\hat{A}^{CSC}$ and the rows of $B$ corresponding to $coarseRowsB$.
    A prefix sum of the resulting histogram generates the coarse-level offsets.
    \item Generate the coarse level by again reading $\hat{A}^{CSC}$, reading the rows of $B$ corresponding to $coarseRowsB$, and writing the result in $colCoarse$ and $valCoarse$ at the positions determined by the offsets.
    As in the fine-level algorithm, we use non-temporal streaming stores for the reordering.
    \item For each coarse-level chunk, apply the fine-level algorithm.
\end{enumerate}

\begin{figure}[htbp]
\centering
\begin{tabular}{c}
\includegraphics[width=0.9\linewidth]{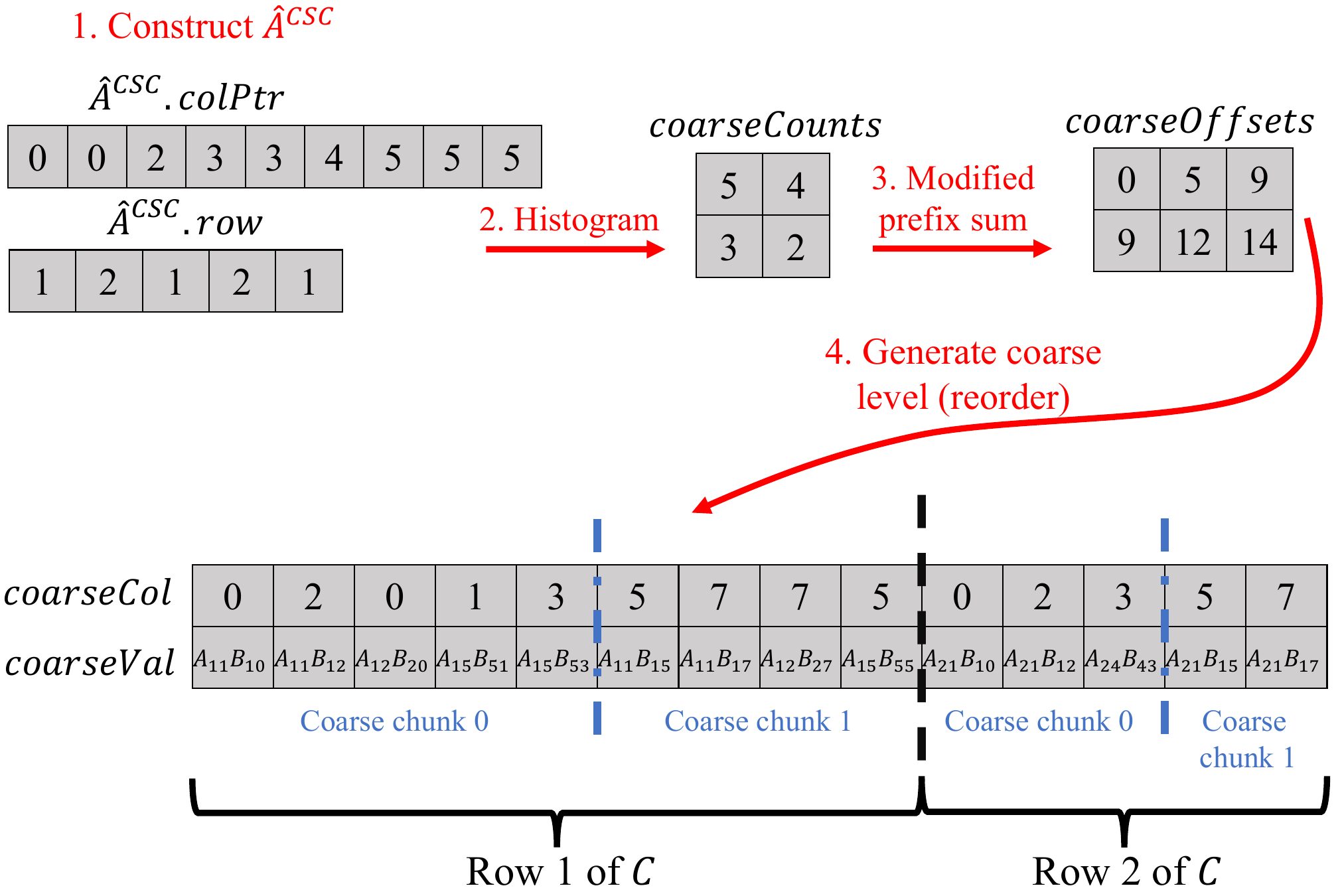} 
\end{tabular}
\caption{Data structure-view of the coarse-level algorithm for the example in \autoref{fig:MAGNUS_example}.}
\label{fig:coarseLevel_example}
\end{figure}

Steps 3-5 are shown in \autoref{alg:magnus_coarse}.
The same set of basic building blocks is used as in the fine-level algorithm (histogramming, prefix summing, and reordering).
The key difference is that the chunks for all rows are generated at the same time.
The prefix sum is modified to compute the chunk offsets within a row based on the offsets from the previous rows.
Just as in the fine-level algorithm, we generate the reordered intermediate product using non-temporal streaming stores.
In the final loop, the fine-level algorithm is applied to each coarse-level chunk for each row.
\autoref{fig:coarseLevel_example} shows the data structures for the example in \autoref{fig:MAGNUS_example}.

\begin{algorithm}[htbp]
    \small
    \caption{MAGNUS coarse-level algorithm}\label{alg:magnus_coarse}
    \KwIn{$\hat{A}^{CSC}$, $\boldsymbol{B}$, $coarseRowsB$, $coarseRowsC$}
    \KwOut{$\boldsymbol{C}_{coarseRowsC}$}
    \SetCommentSty{emph}
    \DontPrintSemicolon
    $countsCoarse \gets 0$\;
    \text{\textbf{\color{blue}/* Histogram*/}}\;
    \For{$i \in coarseRowsB$}{
        \For{$j \gets \hat{A}^{CSC}.colPtr[i]$\textbf{ to }$\hat{A}^{CSC}.colPtr[i+1]-1$}{
            \For{$k \gets \boldsymbol{B.rowPtr}[i]$\textbf{ to }$\boldsymbol{B.rowPtr}[i+1]-1$}{
                $chunk \gets \boldsymbol{B.col}[k]$\texttt{>>}$shiftCoarse$\;
                $countsCoarse[\hat{A}^{CSC}.row[j]][chunk]$\texttt{++}\;
            }
        }
    }
    \text{\textbf{\color{blue}/* Prefix sum */}}\;
    $offsetsCoarse[0][0] \gets 0$\;
    \For{$i \in coarseRowsC$}{
        $\&offsetsCoarse[i][1] \gets$ \texttt{inclusiveScan(}$countsCoarse[i]$\texttt{)}\;
        $offsetsCoarse[i+1][0] \gets offsetsCoarse[i][n_{chunksCoarse}] $\;
    }
    \text{\textbf{\color{blue}/* Reorder */}}\;
    $countsCoarse \gets 0$\;
    \For{$i \in coarseRowsB$}{
        \For{$j \gets \hat{A}^{CSC}.colPtr[i]$\textbf{ to }$\hat{A}^{CSC}.colPtr[i+1]-1$}{
            \For{$k \gets \boldsymbol{B.rowPtr}[i]$\textbf{ to }$\boldsymbol{B.rowPtr}[i+1]-1$}{
                $chunk \gets \boldsymbol{B.col}[k]$\texttt{>>}$shiftCoarse$\;
                $\ell \gets offsetsCoarse[\hat{A}^{CSC}.row[j]][chunk]+countsCoarse[\hat{A}^{CSC}.row[j]][chunk]$\texttt{++}\;
                $colChunks[\ell] \gets \boldsymbol{B.col}[k]-chunk\times chunkLenCoarse$\;
                $valChunks[\ell] \gets \hat{A}^{CSC}.val[j]\times \boldsymbol{B.val}[k]$\;
            }
        }
    }
    \text{\footnotesize\textbf{\color{blue}/* Apply fine-level algorithm to each coarse-level chunk */}}\;
    \For{$i \in coarseRowsC$}{
        \For{$j \gets 0$\textbf{ to }$n_{chunksCoarse}-1$}{
            $k \gets offsetsCoarse[i][j]$\;
            $\boldsymbol{C}_{i,rangeCoarse_j} \gets$ \texttt{fineLevel(}$\&colCoarse[k],\&valCoarse[k]$\texttt{)}\;
        }
    }
\end{algorithm}

The coarse-level algorithm utilizes a set of data structures similar to that of the fine-level algorithm, but without an accumulator. The absence of a cached accumulator enables the coarse-level algorithm to generate more chunks than the fine-level algorithm. However, this comes at the cost of increased data volume due to the additional pass over the intermediate product.
Despite this tradeoff, for a sufficiently large matrix, the increased data volume proves more efficient than the frequent cache misses incurred by the fine-level algorithm, as we will demonstrate in our microbenchmark results.

Similarly to other outer product-based approaches, the memory requirement for the coarse-level algorithm is higher than the fine-level algorithm since the intermediate product for multiple rows must be stored.
In the worst case, the memory requirement is proportional to the sum of the outer products of all rows, which may exceed the memory capacity of certain memory-constrained systems.
Therefore, we use a batching approach, where we collect rows of $C$ into $coarseRowsC$ until one of two conditions is met:
(1) the memory limit of our system is reached, or (2) the storage requirement for generating the coarse-level chunks ($countsCoarse$, $offsetsCoarse$, and the small buffers for the non-temporal streaming stores) exceeds the L2 cache size.
If either condition is met, the batch of rows in $coarseRowsC$ is computed via steps 2-5.
This batching process is repeated until all rows requiring coarse-level locality have been computed.

\subsection{Accumulation}\label{sec:magnus_accum}
MAGNUS is accumulator-agnostic, allowing for portability and flexibility: only the storage requirements of the desired accumulators
are needed to compute the optimal MAGNUS parameters (the optimal parameters are derived in \autoref{sec:magnus_opt_params}.
For portability, accumulators optimized for specific architectures can be selected without changing the locality-generation algorithms.
This is evident in \autoref{alg:magnus_fine}, where the \texttt{accum()} function requires only an input chunk.
For flexibility, MAGNUS allows for a hybrid approach in which each chunk chooses an accumulator based on the chunk characteristics.
In this paper, we consider two accumulators: AVX-512 vectorized bitonic sorting~\cite{AVX512sort} and classical dense accumulation.
For chunks with a small number of elements, sorting is performed. Otherwise, dense accumulation is used.
When visiting each chunk, a threshold is used to choose the accumulator.
This threshold is based on how the sorting algorithm is implemented: quicksort partitions the array, and then hard-coded, vectorized bitonic sorting networks sort the partitions.
We found experimentally that dense accumulation is faster than sorting unless the sort size is small enough to bypass the quicksort algorithm and directly use bitonic sorting.
For more details, see \autoref{sec:results_sort} and~\cite{AVX512sort}.

\subsection{Choosing the Number of Chunks}\label{sec:magnus_opt_params}

The optimal number of chunks is chosen based on the following input parameters, which are readily available for end users: $s_{cacheLine}$, $s_{L2}$, and $m_C$.
Parameters $s_{cacheLine}$ and $s_{L2}$ are the cache line and L2 cache sizes, respectively, which are retrieved by querying the underlying system, e.g., by using standard Linux commands.
Parameter $m_C$ is the number of columns of $C$, which is already included in the CSR data structure of $B$ since $m_C = m_B$.

As explained in \autoref{sec:magnus_fineLevel}, the goal is to retain in the L2 cache certain data structures needed by the fine-level algorithm.
For simplicity, assume $m_C$ is a power of two ($m_C$ is ceiled to the nearest power of two otherwise).
Choosing the optimal number of fine-level chunks corresponds to selecting the value of $n_{chunksFine}$ that minimizes the convex function
\begin{equation}
s_{fineLevel} = \frac{m_C s_{denseAccum}}{\boldsymbol{n_{chunksFine}}} + \boldsymbol{n_{chunksFine}} s_{chunkFine},
\label{equ:fine_level_storage}
\end{equation}
where $s_{fineLevel}$ is the storage requirement for the L2-cached fine-level data structures.
The first term is the storage requirement of the dense accumulator, where the number of elements in the underlying dense accumulator array is $m_C/n_{chunksFine}$. 
For the numeric phase, $s_{denseAccum} = s_{val}+1$, since we need $denseAccumBuff$ and $bitMap$ (see \autoref{alg:gustav_dense}).
For the symbolic phase, $s_{denseAccum} = 1$ since we only need $bitMap$.
The second term is the storage requirement for reordering,
where the storage cost per fine-level chunk is $s_{chunkFine} = s_{histoType}+s_{prefixSumType}+2 s_{cacheLine}$.
The parameters $s_{histoType}$ and $s_{prefixSumType}$ denote the size of the histogram and prefix sum array data types, respectively, which are both four bytes.
The term $2 s_{cacheLine}$ accounts for the storage of active cache lines when noncontiguously writing to $colFine$ and $valFine$ during the reordering phase.
The value of $n_{chunksFine}$ that minimizes \autoref{equ:fine_level_storage} is
\begin{equation}
\boldsymbol{n_{chunksFine}}=\sqrt{\frac{m_C s_{denseAccum}}{s_{chunkFine}}},
\label{equ:fine_level_nchunks_optimal}
\end{equation}
rounded to the nearest power of two, which
is the number of fine-level chunks used 
when the fine-level algorithm is used alone.

By plugging in $n_{chunksFine}$ from \autoref{equ:fine_level_nchunks_optimal} into \autoref{equ:fine_level_storage} we get
\begin{equation}
s_{fineLevel}=2\sqrt{m_C s_{denseAccum} s_{chunkFine}},
\label{equ:fine_level_storage_optimal}
\end{equation}
which is the total storage requirement of the fine-level algorithm when the optimal number of fine-level chunks is used.
When $m_C$ becomes sufficiently large, $s_{fineLevel}$ exceeds the size of the L2 cache, at which point the coarse-level algorithm is applied.
We can now derive the number of fine- and coarse-level chunks when both levels of locality are used.
In this case, we first determine $m_{C_{maxL2}}$, which is the maximum number of columns in which $s_{fineLevel} \leq s_{L2}$.
To calculate $m_{C_{maxL2}}$, we replace $m_C$ with $m_{C_{maxL2}}$ in \autoref{equ:fine_level_storage_optimal} and solve $s_{fineLevel}=s_{L2}$ for $m_{C_{maxL2}}$.
This gives us
\begin{equation}
m_{C_{maxL2}} = \frac{s_{L2}^2}{4  s_{denseAccum}   s_{chunkFine}},
\label{equ:mCmaxL2}
\end{equation}
which is constant and is floored to the nearest power of two.
Therefore,
\begin{equation}
\boldsymbol{n_{chunksFine}}=\sqrt{\frac{m_{C_{maxL2}} s_{denseAccum}}{s_{chunkFine}}},
\label{equ:fine_level_nchunks_optimal2}
\end{equation}
is the optimal number of fine-level chunks when both levels of locality are used, and
the optimal number of coarse-level chunks is
\begin{equation}
\boldsymbol{n_{chunksCoarse}}=m_C / m_{C_{maxL2}}.
\label{equ:coarse_level_nchunks_optimal}
\end{equation}
Equations~\ref{equ:mCmaxL2} and \ref{equ:fine_level_nchunks_optimal2} show that for a sufficiently large value of $m_C$, the number of fine-level chunks stops growing once the coarse-level algorithm is applied.
At this point, the number of coarse-level chunks begins to grow while maintaining a fixed number of fine-level chunks, ensuring that the fine-level data structures fit into the L2 cache.

Note that we only derive optimal parameters for dense accumulation.
This is because the sort-based accumulator does not require any additional storage since
the arrays storing the intermediate product are directly sorted.
The same analysis can also be applied to other accumulators (e.g., hash maps) by modifying \autoref{equ:fine_level_storage}.
\begin{figure}[t]
\centering
\begin{tabular}{c}
\includegraphics[width=\figwidth]{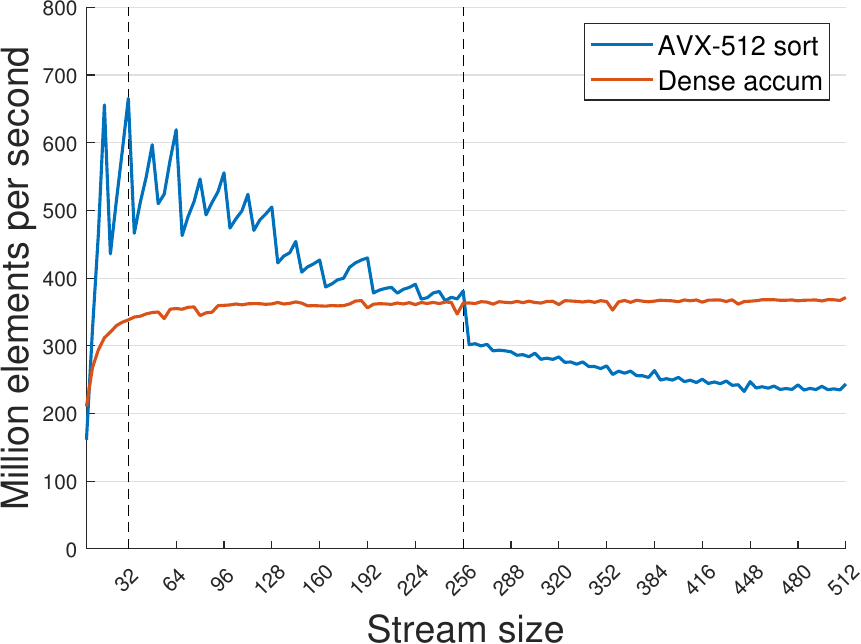}
\end{tabular}
\caption{Comparison of accumulators used in MAGNUS: AVX-512 vectorized sorting and dense accumulation.
A single core of the SPR system is used (see \autoref{tab:machines}).
The rate (millions of elements per second) versus the number of accumulated elements is shown.
The dashed lines indicate where the sorting achieves peak performance (32 elements) and where the performance of dense accumulation overtakes that of sorting (256 elements).}
\label{fig:sort_benchmark}
\end{figure}

\section{Experimental Results}\label{sec:results}

\subsection{Microbenchmarks}\label{sec:microbench}
This section aims to establish our motivation by evaluating the core building blocks of MAGNUS in isolation through microbenchmarking.
The input and output in our microbenchmarks are streams consisting of two arrays of the same size: one of unsigned integers (the \textit{index array}) and the other of floating-point numbers (the \textit{value array}), which emulate the column indices and values, respectively, in the intermediate product.
There are two critical parameters to which SpGEMM accumulators are sensitive: the number of elements in the stream and the maximum value of the elements in the index array (the \textit{stream length}).
The index array is uniformly random in the range $[0, \text{stream length})$, which emulates the range of column indices in the matrix or a MAGNUS chunk.

First, we show that increasing the stream size and length degrades the performance of conventional accumulators.
We then benchmark the building blocks of MAGNUS to demonstrate how the introduction of locality generation can improve the performance of the dense accumulator.
For these experiments, four-byte types were used (\texttt{uint32\_t} and \texttt{float}) on one core of the SPR system (see \autoref{tab:machines}).
We used Likwid~\cite{likwid} to collect performance metrics, providing insight into the cache behavior of the building blocks of MAGNUS.
Likwid is a performance monitoring tool that reports detailed CPU hardware metrics, including the volume of L2-to-L3 cache evictions.

\subsubsection{Accumulators}\label{sec:results_sort}
We first consider the two accumulators used by MAGNUS: AVX-512 vectorized bitonic sorting~\cite{AVX512sort} and dense accumulation.
\autoref{fig:sort_benchmark} shows the rate (in millions of elements per second) versus the stream size for a fixed stream length of $2^{18}$, which is the stream length for which the dense accumulation arrays fit into the L2 cache.
There are two important sizes to consider: 32 and 256. At a size of 32, the sorting achieves peak performance, processing nearly 700 million elements per second. Therefore, targeting sort sizes as close as possible to 32 is ideal.
In MAGNUS, we do precisely this: we combine consecutive chunks until the difference between the sort size and 32 is minimized.
Dense accumulation overtakes sorting at a stream size of 256 and is $\approx 1.5\times$ faster for a sort size of 512.
The 256 threshold originates from the sorting algorithm, which partitions the array into 256 parts (for more information, see~\cite{AVX512sort}).
MAGNUS uses this threshold when selecting an accumulator within a chunk.

\begin{figure}[t]
\centering
\begin{tabular}{c}
\includegraphics[width=\figwidth]{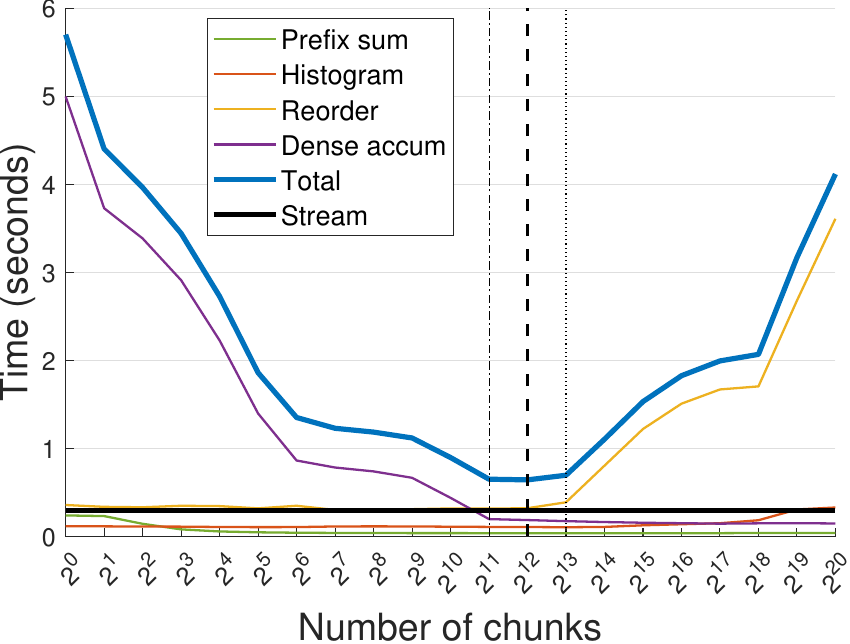} \\
\includegraphics[width=\figwidth]{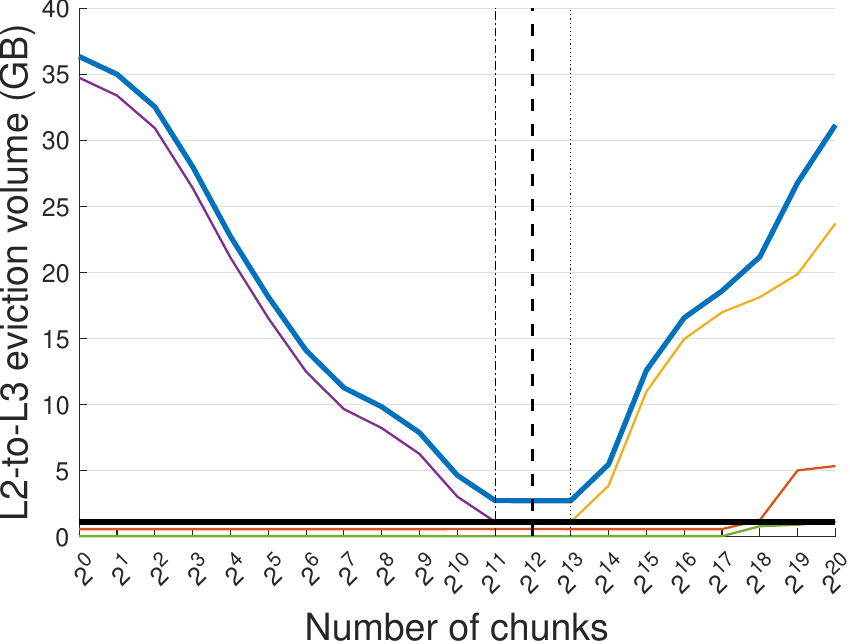}
\end{tabular}
\caption{Wall-clock time (top) and L2-to-L3 cache evictions (bottom) versus the number of chunks for a set of microbenchmarks that test the performance of the building blocks of MAGNUS.
A single core of the SPR system is used (see \autoref{tab:machines}). \textit{Total} denotes the sum of the building block times, and \textit{Stream} is a standard streaming benchmark that serves as the peak-performance baseline. The middle vertical dashed line denotes the optimal number of fine-level chunks.
The left and right dashed lines denote the point at which the storage requirement of the reorder and dense accumulation arrays exceed the L2 cache size, respectively.}
\label{fig:building_blocks_varyChunks}
\end{figure}

\subsubsection{Building Blocks of MAGNUS}
In this experiment, MAGNUS is deconstructed into its building blocks: histogramming, prefix summing, reordering, and accumulation.
For the algorithmic details of each building block, see the associated comments in \autoref{alg:magnus_fine} and \autoref{alg:magnus_coarse}. For example, the algorithm for histogramming is on lines 3-6 in \autoref{alg:magnus_fine} and lines 3-10 in \autoref{alg:magnus_coarse}.
\autoref{fig:building_blocks_varyChunks} shows time and the volume of L2-to-L3 cache evictions (measured using Likwid~\cite{likwid}) versus the number of chunks for a stream size and length of $2^{29}$ elements.
This stream length results in the size of the dense accumulation array varying from $2^{29}$ to $2^{29}/2^{20} = 512$ as the number of chunks increases.
This emulates the per-chunk dense accumulation in MAGNUS.
The time for a standard streaming benchmark is shown, which consists of performing contiguous reads from the input arrays and contiguous writes to the output arrays.
This serves as our peak performance baseline, where the total time (i.e., the sum of all the building block times) cannot exceed the streaming time.
The left and right dashed lines are the points at which the reorder and dense accumulation data structures exceed the L2 cache size.
The middle vertical dashed line shows the optimal number of fine-level chunks (derived in \autoref{sec:magnus_opt_params}), which is calculated using the stream length in place of $m_C$ in \autoref{equ:fine_level_nchunks_optimal}.

The total time, which closely approximates the performance of the fine-level algorithm, is dominated by dense accumulation and reordering.
The time for reordering increases significantly past $2^{13}$, where the storage requirement (active cache lines, histogram array, and prefix sum array) exceeds the L2 cache size, as seen by the increase in L2-to-L3 cache evictions.
For dense accumulation, performance improves as the number of chunks increases due to the reduced size of the dense accumulation array; this is a key result of our locality generation approach.
The total execution time reaches a minimum at the optimal number of fine-level chunks, where both dense accumulation and reordering achieve optimal cache behavior. At this point, the total time is $\approx 2.2$ times the streaming time. Although lower-level optimizations, such as vectorized histogramming, could further decrease this slowdown, maintaining this reasonably small multiple of the peak performance is crucial for scaling to massive matrices.
We will show in \autoref{sec:results_spgemm} that MAGNUS can maintain a similar multiple of the peak performance for massive random matrices, while other SpGEMM baselines cannot.

\begin{figure}[t]
\centering
\begin{tabular}{c}
\includegraphics[width=\figwidth]{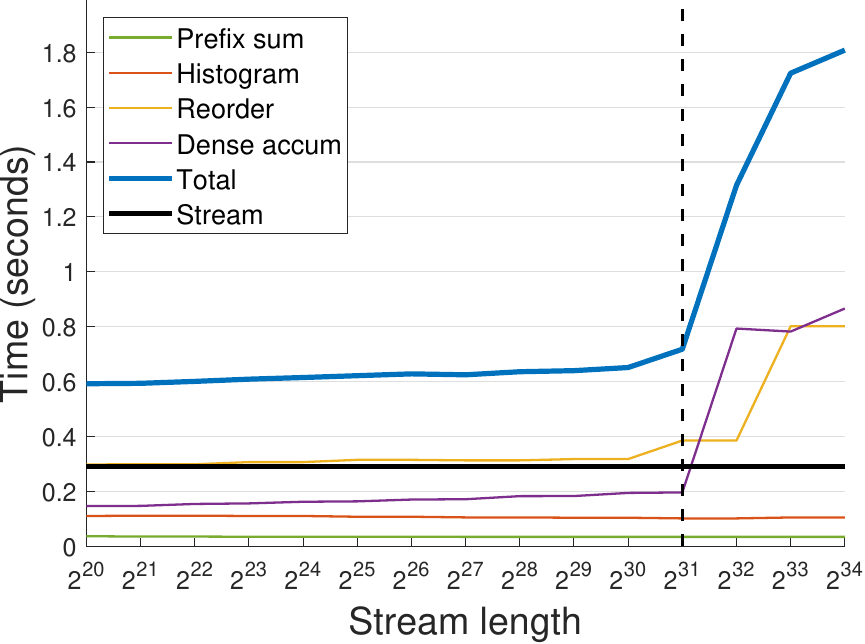} \\
\includegraphics[width=\figwidth]{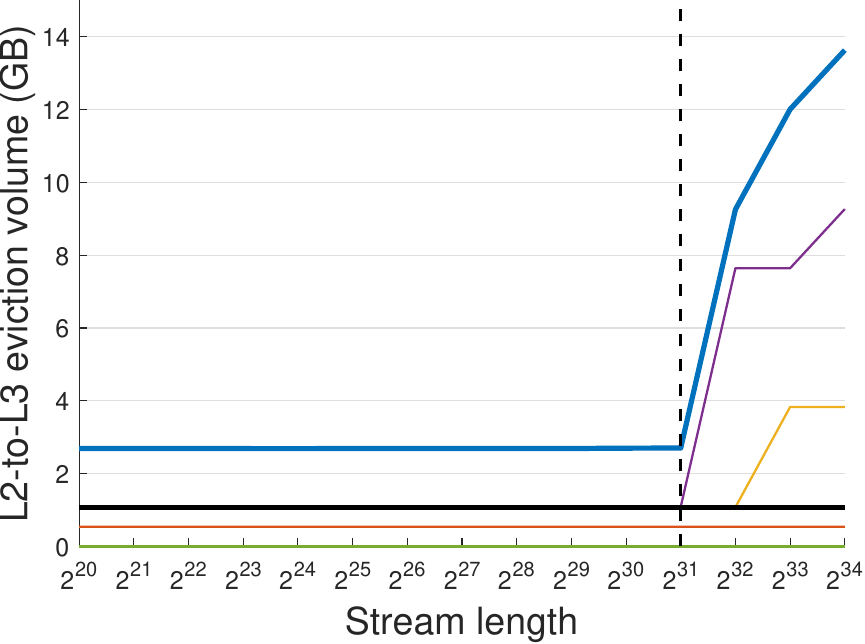}
\end{tabular}
\caption{Wall-clock time (top) and L2-to-L3 cache evictions (bottom) versus the stream length of the input stream for a set of microbenchmarks that test the performance of the building blocks of MAGNUS.  For each stream length value, the optimal number of fine-level chunks is chosen (see \autoref{sec:magnus_opt_params}). \textit{Total} denotes the sum of the building block times, and \textit{Stream} is a standard streaming benchmark that serves as the peak-performance baseline.
The vertical dashed line denotes the threshold at which the fine-level data structures exceed the L2 cache capacity.}
\label{fig:building_blocks_varyMaxIdx}
\end{figure}

\autoref{fig:building_blocks_varyMaxIdx} illustrates the impact of varying the stream length, where the optimal number of fine-level chunks is chosen for each value of the stream length.
Past a stream length of $2^{31}$, the total time rises sharply due to the high volume of L2-to-L3 cache evictions.
This behavior highlights the necessity of the coarse-level algorithm in MAGNUS: even when the optimal number of fine-level chunks is used, the total storage cost of all data structures can exceed the L2 cache capacity when the number of columns of $C$ is sufficiently large.
In MAGNUS, the coarse-level algorithm automatically activates after the breaking point at $2^{31}$.
Consequently, each coarse-level chunk contains column indices in the range $[0, 2^{31})$, allowing the fine-level data structures to fit into the L2 cache.
The fine-level algorithm is then applied to each coarse-level chunk and is cache-efficient.

In summary, our microbenchmarks give us two key conclusions.
First, when both the stream length and size are large, neither sort-based nor dense accumulation performs optimally. However, the relatively inexpensive reordering mechanism in MAGNUS effectively mitigates these issues by reducing both of these quantities. For chunks with a small number of elements, sorting can be applied. Otherwise, dense accumulation is more efficient due to the short stream length per chunk, which allows the accumulation data structures to fit in the L2 cache.

Second, our microbenchmarks provide insight into the overall performance of the coarse- and fine-level algorithms.
The reordering microbenchmark closely approximates the coarse-level algorithm since the dominant cost of the coarse-level algorithm is its reordering phase.
Therefore, the coarse-level algorithm is approximated to perform at near-streaming speed up to the breaking point of $2^{13}$ chunks.
For realistic matrices, this breaking point is likely not reached since that would require the multiplication of matrices with more than $2^{31}2^{13} = 2^{44}$ columns.
The performance of the fine-level algorithm is closely approximated by \textit{Total}, which is dominated by the time for reordering and dense accumulation.
The significant decline in the performance of \textit{Total} beyond a stream length of $2^{31}$ underscores the importance of the initial, near-streaming-speed pass over the intermediate product provided by the coarse-level algorithm.
Rather than using the fine-level algorithm beyond this $2^{31}$ threshold, a comparatively inexpensive initial reordering step yields better performance.
Note that these thresholds ($2^{13}$ and $2^{31}$) are system-dependent; MAGNUS automatically calculates these thresholds using the input system parameters (see \autoref{sec:magnus_opt_params}).

\subsection{Test Configuration for SpGEMM}
\begin{figure*}[!htbp]
\centering
\begin{tabular}{m{0.03\linewidth} m{0.97\linewidth}}
\textbf{EMR} &
\includegraphics[width=0.96\linewidth]{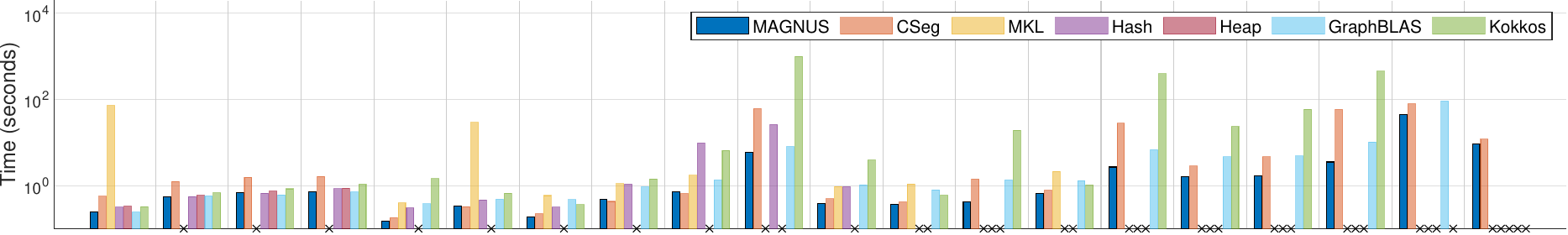} \\
\textbf{SPR} &
\includegraphics[width=0.96\linewidth]{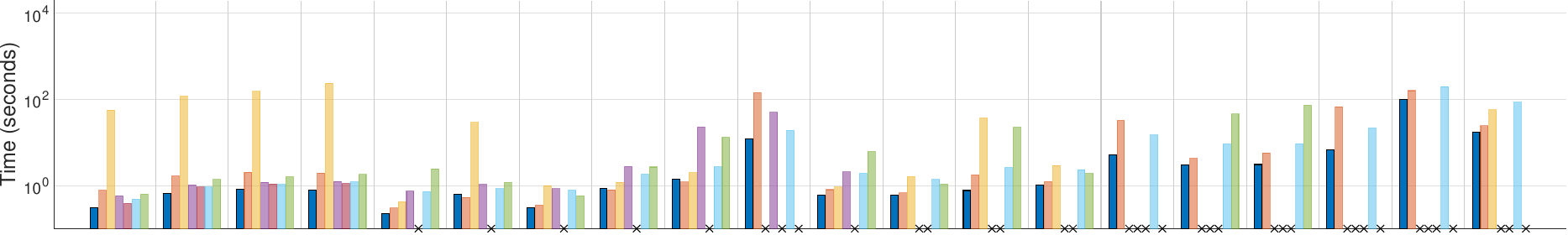} \\
\raisebox{2.5em}{\textbf{SKX}} &
\includegraphics[width=0.96\linewidth]{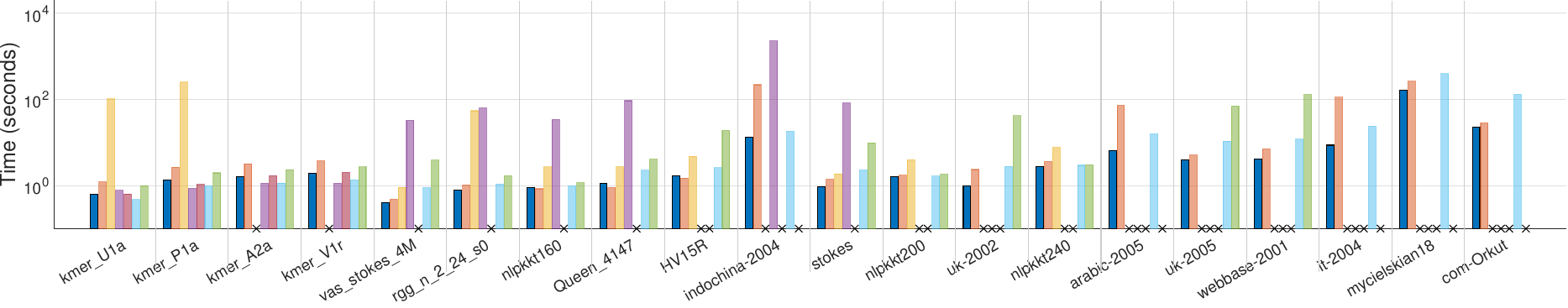} \\
\end{tabular}
\caption{Wall-clock time in log scale for the SuiteSparse matrix collection.
The $\times$-shaped markers denote failed runs (out-of-memory or segmentation faults) of the baselines.
All available threads were used for each system.}
\label{fig:baselines_suitesparse}
\end{figure*}

In the next section, we compare an OpenMP implementation of MAGNUS to a diverse set of state-of-the-art baselines:
CSeg~\cite{cseg}, Intel MKL~\cite{mkl}, vectorized hash/heap-based algorithms~\cite{nagasaka1,nagasaka2}, SuiteSparse:GraphBLAS~\cite{graphblas}, and Kokkos~\cite{kokkos,kokkos2}.
For MKL, we use the sparse BLAS inspector-executor API, i.e., the function \texttt{mkl\_sparse\_spmm()}.
CSeg is the only baseline that implements a locality-generation algorithm, and to the best of our knowledge, it is the only algorithm other than MAGNUS that does so.

We report the total SpGEMM time for all algorithms, where the total time is the sum of the pre-processing, symbolic, numeric, and post-processing phases.
For example, the total time for CSeg is the sum of the time taken to construct the high-level summary matrix and perform the symbolic and numeric phases.  
In contrast, for MKL, we measure only the time of the call to \texttt{mkl\_sparse\_spmm()}, as that is the only operation exposed to us.
For MAGNUS, the total time is the sum of the setup, symbolic, and numeric phases.
We perform one warm-up run and then extract the time by taking the average of the next 10 runs.
We found that the time did not vary significantly between many runs.
Our test systems are shown in \autoref{tab:machines}, all of which use Intel processors.
We utilized all available threads, including hyperthreads, for all SpGEMM runs, resulting in the fastest execution time for both the baseline implementations and MAGNUS.

\begin{table}[htbp]
\caption{Hardware specifications of the test systems.
All systems are a single multisocket node with Intel CPUs.}
\begin{center}
\resizebox{\columnwidth}{!}{
\begin{tabular}{lrrr}
\hline
Architecture & Skylake (SKX) & Sapphire Rapids (SPR) & Emerald Rapids (EMR) \\
\hline
Xeon Model & Gold 6140 & Gold 6438M & Platinum 8592+ \\
Sockets & 4 & 2 & 2 \\
Total cores and threads & 72 and 144 & 64 and 128 & 128 and 256 \\
L1 size per core & 32 KB & 48 KB & 48 KB \\
L2 size per core & 1 MB & 2 MB & 2 MB \\
L3 size per socket & 24.75 MB & 60 MB & 320 MB \\
Memory & 2 TB & 4 TB & 1 TB \\
\hline
\end{tabular}
}
\label{tab:machines}
\end{center}
\end{table}

\begin{table}[htbp]
\caption{Properties of the SuiteSparse matrices.}
\begin{center}
\resizebox{\columnwidth}{!}{
\begin{tabular}{lrrrrr}
\hline
Matrix & $n_A$ & $nnz_A$ & $nnz_{A^2}/n_{A^2}$ & $nnz_{A^2}$ \\
\hline
kmer\_U1a & 67,716,231 & 138,778,562 & 3.3 & 222,262,359 \\
kmer\_P1a & 139,353,211 & 297,829,984 & 3.8 & 531,367,449 \\
kmer\_A2a & 170,728,175 & 360,585,172 & 3.6 & 622,660,207 \\
kmer\_V1r & 214,005,017 & 465,410,904 & 3.9 & 824,450,881 \\
vas\_stokes\_4M & 4,382,246 & 131,577,616 & 188.6 & 826,486,449 \\
rgg\_n\_2\_24\_s0 & 16,777,216 & 265,114,400 & 49.4 & 828,639,073 \\
nlpkkt160 & 8,345,600 & 229,518,112 & 148.7 & 1,241,294,184 \\
Queen\_4147 & 4,147,110 & 329,499,284 & 362.2 & 1,501,950,816 \\
HV15R & 2,017,169 & 283,073,458 & 876.5 & 1,768,066,720 \\
indochina-2004 & 7,414,866 & 194,109,311 & 263.3 & 1,952,630,542 \\
stokes & 11,449,533 & 349,321,980 & 184.7 & 2,115,146,825 \\
nlpkkt200 & 16,240,000 & 448,225,632 & 149.4 & 2,425,937,704 \\
uk-2002 & 18,520,486 & 298,113,762 & 172.5 & 3,194,986,138 \\
nlpkkt240 & 27,993,600 & 774,472,352 & 149.8 & 4,193,781,224 \\
arabic-2005 & 22,744,080 & 639,999,458 & 366.0 & 8,323,612,632 \\
uk-2005 & 39,459,925 & 936,364,282 & 227.4 & 8,972,400,198 \\
webbase-2001 & 118,142,155 & 1,019,903,190 & 114.0 & 13,466,717,166 \\
it-2004 & 41,291,594 & 1,150,725,436 & 340.2 & 14,045,664,641 \\
mycielskian18 & 196,607 & 300,933,832 & 195,076.4 & 38,353,378,617 \\
com-Orkut & 3,072,441 & 234,370,166 & 16,220.6 & 49,836,711,933 \\
\hline
\end{tabular}
}
\label{tab:suite_sparse}
\end{center}
\end{table}

\begin{table}[htbp]
\caption{Properties of the RMat16 matrices (R-mats with an average of 16 nonzero entries per row).
The standard Graph500 parameters are used ($a = .57$ and $b = c = .19$).}
\begin{center}
\resizebox{\columnwidth}{!}{
\begin{tabular}{lrrrrr}
\hline
Scale & $n_A$ & $nnz_A$ & $nnz_{A^2}/n_{A^2}$ & $nnz_{A^2}$ \\
\hline
18 & 262,144 & 4,194,304 & 5,141.7 & 1,347,858,618 \\
19 & 524,288 & 8,388,608 & 7,072.6 & 3,708,083,907 \\
20 & 1,048,576 & 16,777,216 & 9,479.9 & 9,940,402,266 \\
21 & 2,097,152 & 33,554,432 & 12,377.9 & 25,958,392,028 \\
22 & 4,194,304 & 67,108,864 & 16,387.1 & 68,732,382,095 \\
23 & 8,388,608 & 134,217,728 & 22,253.2 & 186,673,674,064 \\
\hline
\end{tabular}
}
\label{tab:rmat16_stats}
\end{center}
\end{table}

We consider three important matrix test sets: matrices from the SuiteSparse collection~\cite{suitesparse}, recursive model power-law matrices (R-mats)~\cite{rmat}, and uniform random matrices (i.e., those from the Erd\H{o}s--R\'enyi (ER) model)~\cite{erdosrenyi}.
For the SuiteSparse and R-mat test sets, we consider the operation $A^2$ for square $A$, which is the de facto standard for evaluating SpGEMM algorithms.
The configuration for the uniform random matrix set will be discussed later in this section.
\autoref{tab:suite_sparse} shows the set of SuiteSparse matrices used in our experiments, where $nnz_{A^2}/n_{A^2}$ is the average number of nonzero entries per row of $A^2$, which is a measure of the sparsity of $A^2$.
These matrices are the largest 20 (in terms of the total number of nonzero entries in $A$) in which both the $A$ and $A^2$ (the result of SpGEMM in our experiments) fit into memory for MAGNUS and at least one baseline.
\autoref{tab:rmat16_stats} shows the R-mats with an average of 16 nonzero entries per row, where the table is organized by increasing \emph{scale}, e.g., the scale-18 R-mat has $2^{18}$ rows.
The standard Graph500 parameters were used to generate these R-mats ($a = .57$ and $b = c = .19$).
The scale-23 matrix is the largest in which both the input
and output fit into memory for MAGNUS and at least one baseline on the SPR system.

\subsection{SpGEMM Results}\label{sec:results_spgemm}

\begin{figure}[htbp]
\newcommand{\figwidthLoc}{.95\linewidth}
\centering
\begin{tabular}{c}
\includegraphics[width=\figwidth]{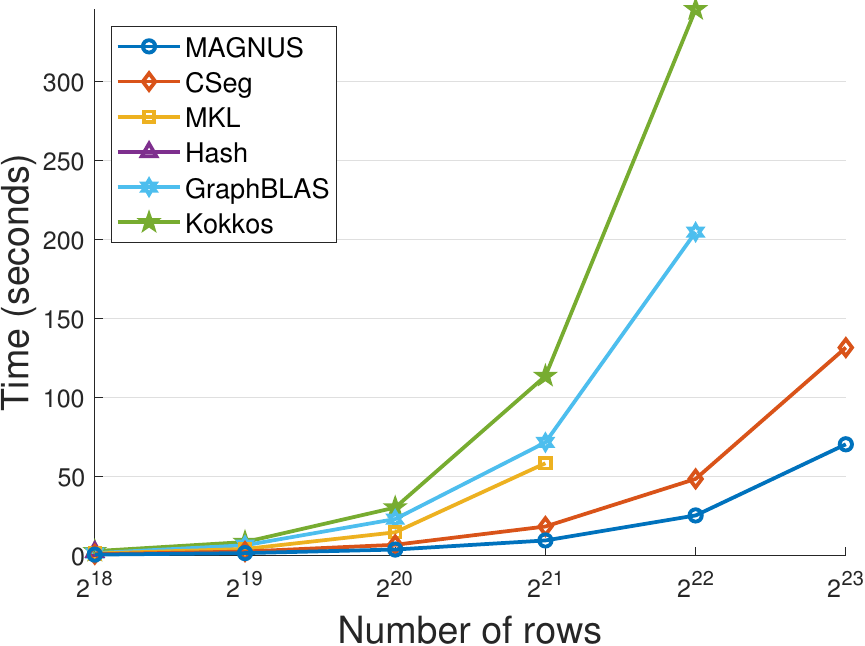} \\
\includegraphics[width=\figwidth]{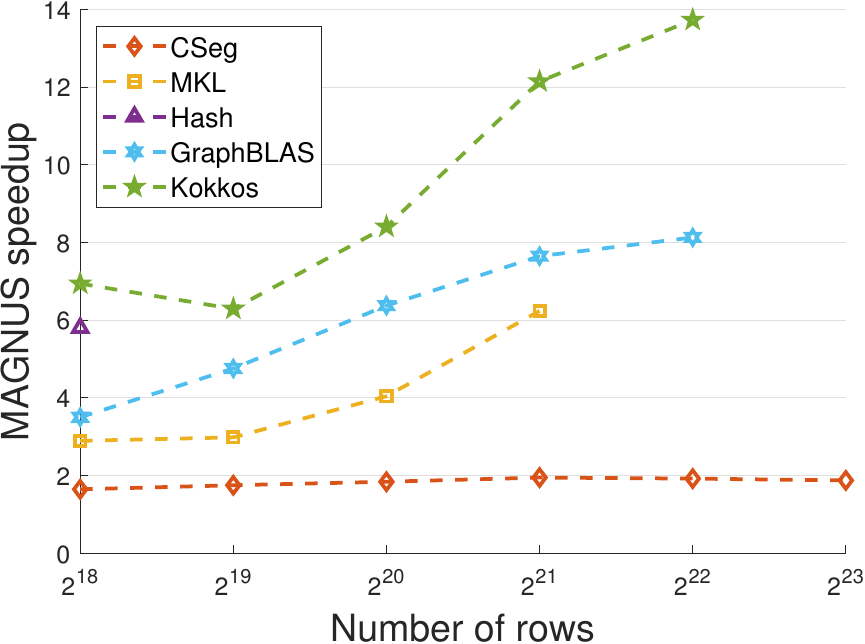}
\end{tabular}
\caption{Wall-clock time and speedup versus number of rows for the RMat16 matrix set on the SPR system (see \autoref{tab:machines}).
The speedup is the ratio of the time of the baselines to that of MAGNUS.}
\label{fig:baselines_rmat16}
\end{figure}

\begin{figure*}[htbp]
\newcommand{\figwidthLoc}{.31\linewidth}
\centering
\begin{tabular}{ccc}
\textbf{EMR} & \textbf{SPR} & \textbf{SKX} \\
\includegraphics[width=\figwidthLoc]{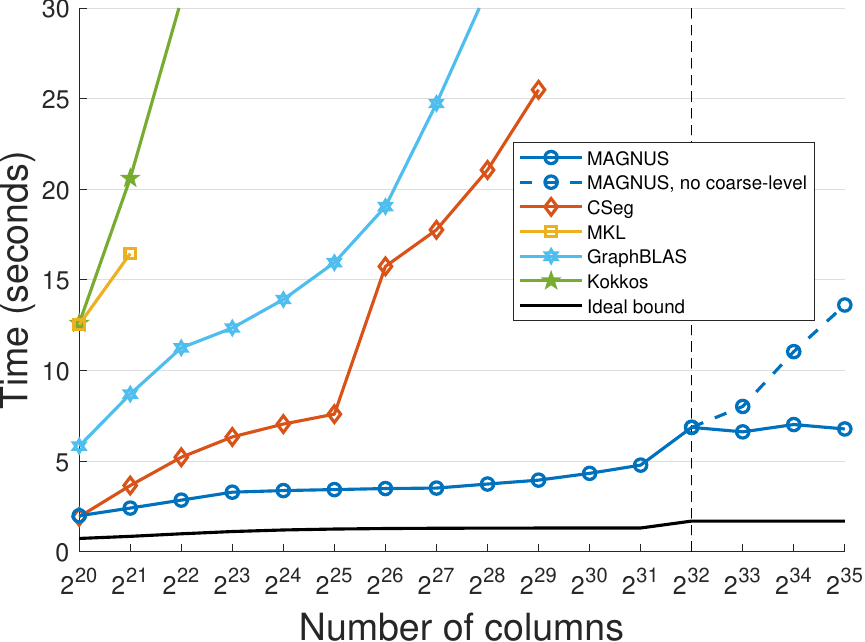} &
\includegraphics[width=\figwidthLoc]{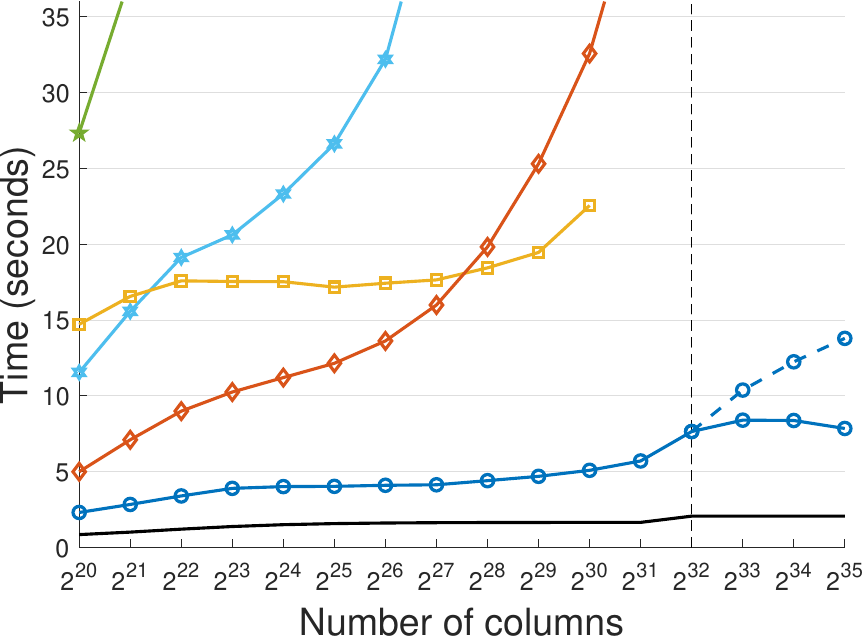} &
\includegraphics[width=\figwidthLoc]{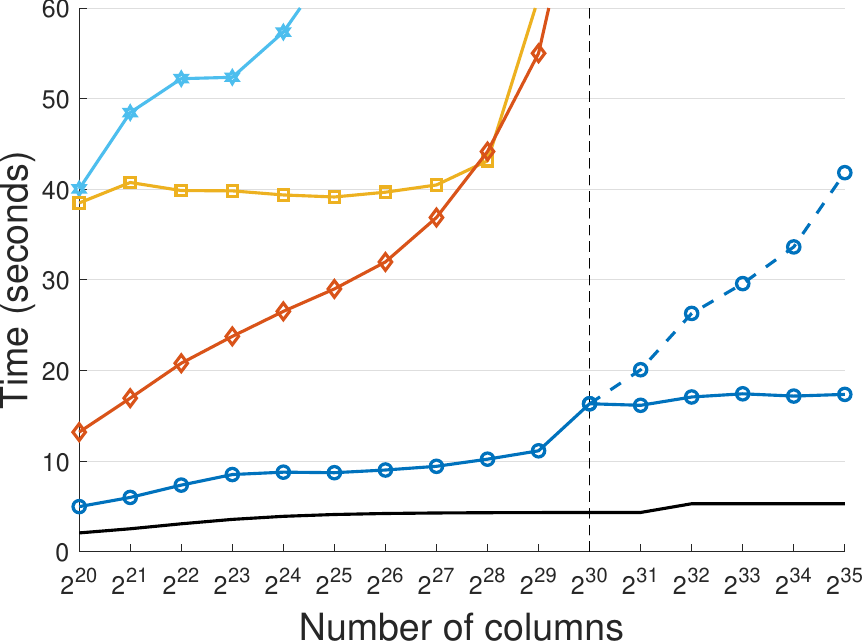}
\end{tabular} 
\caption{Time versus number of columns of $C$ for the uniform random matrix test set.
A fixed average number of nonzero entries per row of 2048 is used.
The vertical dashed line denotes the point at which MAGNUS begins to use the coarse-level algorithm.}
\label{fig:massiveERs_varyScale}
\end{figure*}

We first consider the SuiteSparse matrix collection.
\autoref{fig:baselines_suitesparse} shows the execution time in logarithmic scale for each matrix and each system.
In some cases, out-of-memory errors and segmentation faults caused some of the baselines to fail, denoted by the missing bars.
MAGNUS is often faster than all baselines and is often orders of magnitude faster than at least one baseline.
CSeg is sometimes slightly faster than MAGNUS (up to $1.23\times$) for Queen\_4147, HV15R, rgg\_n\_2\_24\_s0, and nlpkkt160.
For these matrices, MAGNUS places all rows into the dense accumulation category due to their banded structure, i.e., these matrices do not require locality generation to be efficiently multiplied.
This suggests that in the absence of locality generation, the accumulators in CSeg may be slightly more optimized for banded matrices.
Similarly, the kmer matrices also do not require locality generation, where Hash, Heap, and GraphBLAS are slightly faster than MAGNUS in several cases.
These matrices are not banded but are highly sparse, leading to an intermediate product per row that is less than our dense accumulation threshold.
Therefore, sort- or hash map-based accumulators are most effective, as shown by MAGNUS, Hash, and Heap having the fastest times (most rows in MAGNUS are placed into the sort category).

In all other cases, MAGNUS categorizes rows as a mixture of sorting, dense accumulation, and fine-level locality, 
where a significant number of rows require the fine-level algorithm (for all SuiteSparse matrices, the coarse-level algorithm is not needed).
MAGNUS is the fastest method for 76\% of the 60 test cases (20 matrices across three systems) and is always the fastest for the 11 largest matrices.
MAGNUS is up to $16.6$, $306.7$, $172.5$, $1.4$, $5.7$, and $171.8$ times faster than CSeg, MKL, Hash, Heap, GraphBLAS, and Kokkos, respectively, and is only $1.4$ times slower than any of the baselines in the worst case.
The low peak speedup over Heap is due to the high failure rate of Heap, which only ran to completion for the kmer matrices.

\autoref{fig:baselines_rmat16} shows the time and speedup versus the number of rows for the RMat16 matrix set on the SPR system, where the speedup is the ratio of the time of the baselines to that of MAGNUS.
The SPR system, which has the largest memory, allows us to scale to the largest RMat16 matrix.
Showing results for the other two systems does not provide additional insight.
Since we are using the standard Graph500 parameters, a small average number of nonzero entries per row (16 in this case) produces a high amount of fill-in in $C$ (as shown in \autoref{tab:rmat16_stats}) because many of the nonzero entries in $A$ are clustered in the top left corner.
Unlike banded and other highly sparse structures that produce less fill-in and more regular access patterns, the mixture of a random distribution with clustering in the RMat16 matrices is problematic for conventional accumulators since a high data volume across a wide range of column indices results in data structures that do not fit into the L2 cache.
This is demonstrated in \autoref{fig:baselines_rmat16}, which illustrates the poor scaling of all baselines, except for CSeg, as the scale increases.
Heap failed in all cases and Hash failed in all but the smallest case.
MAGNUS is $6.2$, $5.8$, $8.1$, and $13.7$ times faster than MKL, Hash, GraphBLAS, and Kokkos, respectively, for the largest matrices.
The scaling of CSeg and MAGNUS demonstrates the importance of locality generation.
Although MAGNUS is $\approx1.8$ times faster than CSeg, they both scale at a similar rate.

Lastly, we consider uniform random matrices.
Unlike the R-mats, a small number of nonzero entries per row does not produce a high amount of fill-in in $C$. 
However, as we scale up the number of columns and nonzero entries per row of $B$, the uniform distribution of the column indices results in frequent accesses to the entire accumulator.
For conventional accumulators, this becomes cache-inefficient if no locality generation strategy is used.
Since the performance of conventional accumulators is sensitive to the number of columns of $C$ and not to the number of rows (see \autoref{sec:microbench}), we consider the nonsquare case where $C$ has 4096 rows and a variable number of columns.
This allows us to scale to massive matrices without exceeding the memory limit of our systems.
Furthermore, only the rows of $B$ that depend on the nonzero entries in $A$ are generated, saving additional memory.

\autoref{fig:massiveERs_varyScale} shows results for increasing the number of columns with a fixed average number of nonzero entries per row of 2048.
Hash and Heap failed in all cases.
The black line is the ideal performance bound, which is calculated by dividing the minimum required data volume for Gustavson-based SpGEMM by the system bandwidth, i.e., $T_{ideal} = \frac{n_{readVol}+n_{writeVol}}{r_{bw}}$,
where the system bandwidth $r_{bw}$ was measured using a streaming microbenchmark.
The ideal read volume is
\begin{equation}
\begin{aligned}
n_{readVol} = & 2(n_A+1) s_{rowPtr} + nnz_A(4s_{rowPtr} + 2s_{colIdx} + s_{val}) + \\
& n_{interProd} (2s_{colIdx} + s_{val}),
\end{aligned}
\label{equ:ideal_data_vol}
\end{equation}
where $s_{rowPtr}$ is the size of the CSR row pointer type (\texttt{size\_t} in our implementation), $s_{colIdx}$ is the size of the column index type (\texttt{uint32\_t} or \texttt{uint64\_t} depending on the size of the matrix), and $s_{val}$ is the matrix coefficient value type (\texttt{float} in our experiments).
The first term denotes the read volume of the row pointers of $A$; the second term denotes the read volume of the nonzero entries in $A$ and the row pointers in $B$; and the third term, which is the asymptotically dominant term, denotes the read volume of the nonzero entries in the rows of $B$.
The number of elements in the intermediate product is defined as $n_{interProd} = \sum_{(i,j)\in \mathcal{S}(A)}nnz_{B_j}$.
The factors of two account for the symbolic and numeric phases.
The factor of four accounts for the symbolic phase, the numeric phase, and the need to read both the start and end row pointers for each row of $B$ (in contrast to the reading of rows of $A$, the rows of $B$ are read nonconsecutively).

The ideal write volume is 
\begin{equation}
n_{writeVol} = (n_C+1) s_{rowPtr} + nnz_C (s_{colIdx} + s_{val}),
\end{equation}
with the first term corresponding to writing the row pointers of $C$ and the second term to writing the nonzero entries.
The overall data volume is the sum of the read and write data volumes.
This ideal bound does not account for various costs, such as NUMA effects, synchronization overheads, or the performance of accessing intermediate data structures (e.g., the accumulator), which can have a significant impact on the performance of SpGEMM algorithms.
Additionally, this bound does not consider cached rows of $B$, as reflected in the expression for $n_{interProd}$, which assumes that previously read rows of $B$ are not reused.
For this reason, we only consider this bound for the uniform random matrices, where there is minimal opportunity for row reuse in $B$.

\autoref{fig:massiveERs_varyScale} shows that MAGNUS maintains an average multiple of $\approx 2.7$ of the ideal bound before applying the coarse-level algorithm and $\approx 3.5$ afterward.
The increase in the multiple is due to the increase in data volume incurred by the outer product (additionally, the slight increase in the ideal bound at $2^{32}$ is due to a change from \texttt{uint32\_t} to \texttt{uint64\_t} for $s_{colIdx}$).
The $2.7$ multiple is consistent with the multiple from our microbenchmarks.
In contrast, the baselines, including CSeg, diverge from the ideal bound.
This suggests that the locality generation method in CSeg, which explicitly constructs an auxiliary segmented matrix, does not scale if a large number of segments is required.
The vertical dashed line shows the point at which MAGNUS starts to place rows in the coarse-level category.
For SKX, the crossover point occurs at $2^{30}$, compared to $2^{32}$ for SPR and EMR, due to the smaller L2 cache size in SKX.
The dashed blue line shows MAGNUS with the coarse-level algorithm turned off, where MAGNUS diverges from the ideal bound.
This shows the necessity of multiple levels of locality, especially for massive matrices where the fine-level data structures do not fit into the L2 cache.
\section{Conclusion}
On modern CPUs, 
current SpGEMM algorithms often scale poorly to massive
matrices due to inefficient use of the cache hierarchy.
We present MAGNUS, a novel algorithm for locality generation, where
the intermediate product is reordered into cache-friendly chunks using a hierarchical two-level approach.
MAGNUS consists of two algorithms that create multiple levels of locality: the fine- and coarse-level algorithms.
The coarse-level algorithm generates a set of coarse-level chunks, and the fine-level algorithm further subdivides the coarse-level chunks into cache-friendly fine-level chunks.
An accumulator is then applied to each fine-level chunk, where a dense or sort-based accumulator is selected based on a threshold on the number of elements in the chunk.
MAGNUS is input- and system-aware: the chunk properties are determined using the matrix dimensions and the system cache sizes.

Our experimental results compare MAGNUS with several state-of-the-art baselines for three matrix sets on three Intel architectures.
MAGNUS is faster than all the baselines in most cases and is often an order of magnitude faster than at least one baseline.
More importantly, MAGNUS scales to massive, uniform random matrices, the most challenging test sets that we consider.
This challenging case highlights the importance of the locality-generation techniques in MAGNUS, which allows MAGNUS to scale with an ideal performance bound independent of the matrix properties.  In contrast, the baselines diverge from this bound as the matrix size increases.

\bibliographystyle{plain}
\bibliography{refs}

\end{document}